\documentclass[aps,pra,twocolumn, 10pt, showpacs,superscriptaddress,groupedaddress,floatfix]{revtex4-1} 
\usepackage{xspace}
\usepackage[hidelinks]{hyperref}

\newcommand{\pd}[2]{\ensuremath{\frac{\partial #1}{\partial #2}}}

\newcommand{\na}{\ensuremath{\mathbf\nabla}}

\newcommand{\ie}{\emph{i.e.} }
\newcommand{\eg}{\emph{e.g.}\@\xspace }

\newcommand{\NTen}{four}
\newcommand{\NFifteen}{five}
\newcommand{\NTwenty}{seven}

\usepackage{graphicx}
\usepackage{epstopdf}
\usepackage{setspace}
\usepackage{bbold}
\usepackage{comment}
\usepackage{color}
\usepackage{soul}
\usepackage[abs]{overpic}
\usepackage{amsmath}
\usepackage{lipsum}
\usepackage[abs]{overpic}
\usepackage{amsmath}
\usepackage{nicefrac}

\begin{document}

\author{Gadi Afek}
\affiliation{Wright Laboratory, Department of Physics, Yale University, New Haven, Connecticut 06520, USA}
\author{Fernando Monteiro}
\affiliation{Wright Laboratory, Department of Physics, Yale University, New Haven, Connecticut 06520, USA}
\author{Jiaxiang Wang}
\affiliation{Wright Laboratory, Department of Physics, Yale University, New Haven, Connecticut 06520, USA}
\author{Benjamin Siegel}
\affiliation{Wright Laboratory, Department of Physics, Yale University, New Haven, Connecticut 06520, USA}
\author{Sumita Ghosh}
\affiliation{Wright Laboratory, Department of Physics, Yale University, New Haven, Connecticut 06520, USA}
\author{David C. Moore}
\affiliation{Wright Laboratory, Department of Physics, Yale University, New Haven, Connecticut 06520, USA}

\title[]{Limits on the abundance of millicharged particles bound to matter}

\begin{abstract}
Millicharged particles (mCPs) are hypothesized particles possessing an electric charge that is a fraction of the charge of the electron. We report a search for mCPs with charges $\gtrsim 10^{-4}~e$ that improves sensitivity to their abundance in matter by roughly two orders of magnitude relative to previous searches. This search is sensitive to such particles over a wide range of masses and charges for which they can form stable bound states with matter, corresponding to a gap in parameter space that is beyond the reach of previous searches from accelerators, colliders, cosmic-ray experiments, and cosmological constraints. 
\end{abstract}

\maketitle
The discrete nature of electric charge has been a fundamental concept since the measurement of the electron charge, $e$, in the Millikan experiment~\cite{PhysRev.2.109_millikan}. All subsequent measurements for known fundamental particles are consistent with exact quantization of charges in multiples of $e$ (or $e/3$ for quarks)~\cite{Unnikrishnan_2004}. It is, however, possible that particles exist with an electric charge that is a fraction of the charge of the electron, $\vert Q\vert = \varepsilon e$. Typically called ``millicharged particles'' (mCPs)~\cite{PhysRevLett.65.679_mcp,Jaeckel:2010ni_review}, such particles can appear in ``hidden sector'' models for dark matter (DM)~\cite{Holdom1986,PhysRevD.92.035014_mCP_hypercharge_window,PhysRevD.41.1067_old_dark_photon,Pospelov:2007mp,PhysRevD.80.075018_accel_dark_forces,PhysRevD.88.114015_gordan_accel}, where some fraction, $f_Q$, of the local relic density of dark sector particles may consist of mCPs. In addition to searching for mCPs, searches for a deviation in the sum of the proton, electron, and neutron charge from zero can serve as a precise test of the Standard Model (SM)~\cite{Foot_1993} and a probe of beyond SM physics~\cite{LANGACKER1981185_GUT_review,PhysRevD.49.3617_foot_anomaly,Foot_1993}.

Levitated optomechanical sensors have enabled significant advances in precision measurements over the past few years~\cite{2020RPPh...83b6401M_millen_review,Ahn2020,Moore:2020awi,lewandowski2020high,Vinante2020}. SiO$_2$ nanospheres have been cooled to just a few vibrational quanta en-route to testing fundamental concepts related to quantum mechanics~\cite{Tebbenjohanns2019,Kamba2020,Delic2020, Carney2020}, micron-sized spheres cooled to $\sim50~\mu$K have enabled searches for recoils associated with passing dark-matter~\cite{Monteiro2020,Monteiro2020DM}, and atomic systems have allowed, among many other examples, precise measurement of fundamental constants and searches for DM~\cite{Safronova2018,Parker2018,Morel2020,Kennedy2020}.

In this Article we use recently developed ng-mass levitated optomechanical sensors to extend searches for mCPs bound in matter to a total mass of $\sim75$~ng, increasing the abundance sensitivity by two orders of magnitude relative to previous searches~\cite{Moore2014}, while maintaining sensitivity to single mCPs with $\vert\varepsilon\vert\gtrsim 10^{-4}$. This advance is enabled by new techniques to optically levitate objects in high vacuum more than 100$\times$ more massive than previously possible~\cite{Monteiro2017}, the ability to optically rotate these objects at MHz frequencies~\cite{Monteiro2018}, and substantial reductions in technical sources of noise, enabling aN-level force sensitivity for ng mass objects~\cite{Monteiro2020}. For mCPs bound in matter at an average abundance $\gtrsim10^{-17}-10^{-15}$ per nucleon, these results probe a substantial gap in parameter space beyond the reach of previous searches. In addition, the sum of the proton, neutron, and electron charge is constrained to be $\lesssim 3 \times 10^{-19}~e$. While not yet competitive with the best existing constraints~\cite{Bressi2011,Baumann1988}, further suppression of the backgrounds identified here may allow new levitated optomechanical tests of the neutrality of matter.

Previous searches for mCPs include collider~\cite{Davidson2000LEP,Jaeckel2013LHC,CMS:2012xi,Chatrchyan:2013oca}, fixed target~\cite{Acciarri2020ArgoNeuT,Magill:2018tbb}, and dedicated~\cite{Prinz1998SLAC,Ball2020MilliquanDemo} experiments. Detectors for neutrinos~\cite{Plestid:2020kdm,Alvis:2018yte,Singh:2018von} and DM~\cite{alkhatib2020constraints,Agnese:2014vxh} have further constrained the flux of mCPs passing through terrestrial detectors. However, constraints from such searches typically become weaker for mCP masses $\gtrsim 1$~GeV and fractional charges $\lesssim 0.1~e$~\cite{Plestid:2020kdm,Magill:2018tbb,pospelov2020earthbound}. Additionally, existing searches are not expected to be sensitive to relic DM particles in the galactic halo with $v/c \sim 10^{-3}$~\cite{alkhatib2020constraints}, although techniques have been proposed to accelerate them to detectable energies~\cite{pospelov2020earthbound}. 

mCPs may also form bound-states with matter through electrostatic interactions~\cite{Langacker2011,pospelov2020earthbound}, for example forming a Bohr-like ``atom'' consisting of an atomic nucleus and negatively charged mCP. In contrast to the searches described above, it may be possible to detect relic mCPs over a wide mass range by searching for mCPs bound in terrestrial matter, even if they make up only a fraction $f_Q \ll 1$ of the relic density. Recent work to estimate the abundance of relic DM mCPs on earth indicates that their number density can be significantly enhanced relative to their galactic abundance, due to thermalization (and eventual capture) in terrestrial material~\cite{pospelov2020earthbound}. While a detailed calculation of the expected concentration of relic DM mCPs bound in a given terrestrial material is complicated, estimates of the average terrestrial concentration indicate abundances per nucleon as high as $10^{-10}f_Q$ for GeV--TeV scale masses are possible. For mCPs in the region of parameter space in which bound states can be formed (masses $\gtrsim1$ GeV for $\varepsilon\gtrsim10^{-5}$), it is plausible that surface and atmospheric materials may contain even larger abundances of mCPs than the terrestrial average since particles can be captured before sinking to sub-surface depths. 

Previous searches for fractionally charged particles bound in magnetically levitated masses~\cite{LaRue1981,Marinelli1982,Phillips1988,Smith1989} or Millikan-type oil-drop experiments~\cite{Joyce1983,VanPolen1987,Lee2002,Kim2007}, have excluded their presence with $\vert\varepsilon\vert\gtrsim 0.1$, probing total masses of $>100$~mg. An initial search using 5~$\mu$m diameter SiO$_2$ microspheres extended sensitivity to single mCPs with charges as small as $\vert\varepsilon\vert \sim 5 \times 10^{-5}$, but tested only $\sim 1$~ng of mass. By extending these techniques to substantially larger SiO$_2$ spheres (up to 20~$\mu$m in diameter), the results presented here are able to probe roughly 100$\times$ lower abundances throughout the range in which they can form stable bound states. 

\begin{figure} 
  \centering
    \begin{overpic}[width=\linewidth]{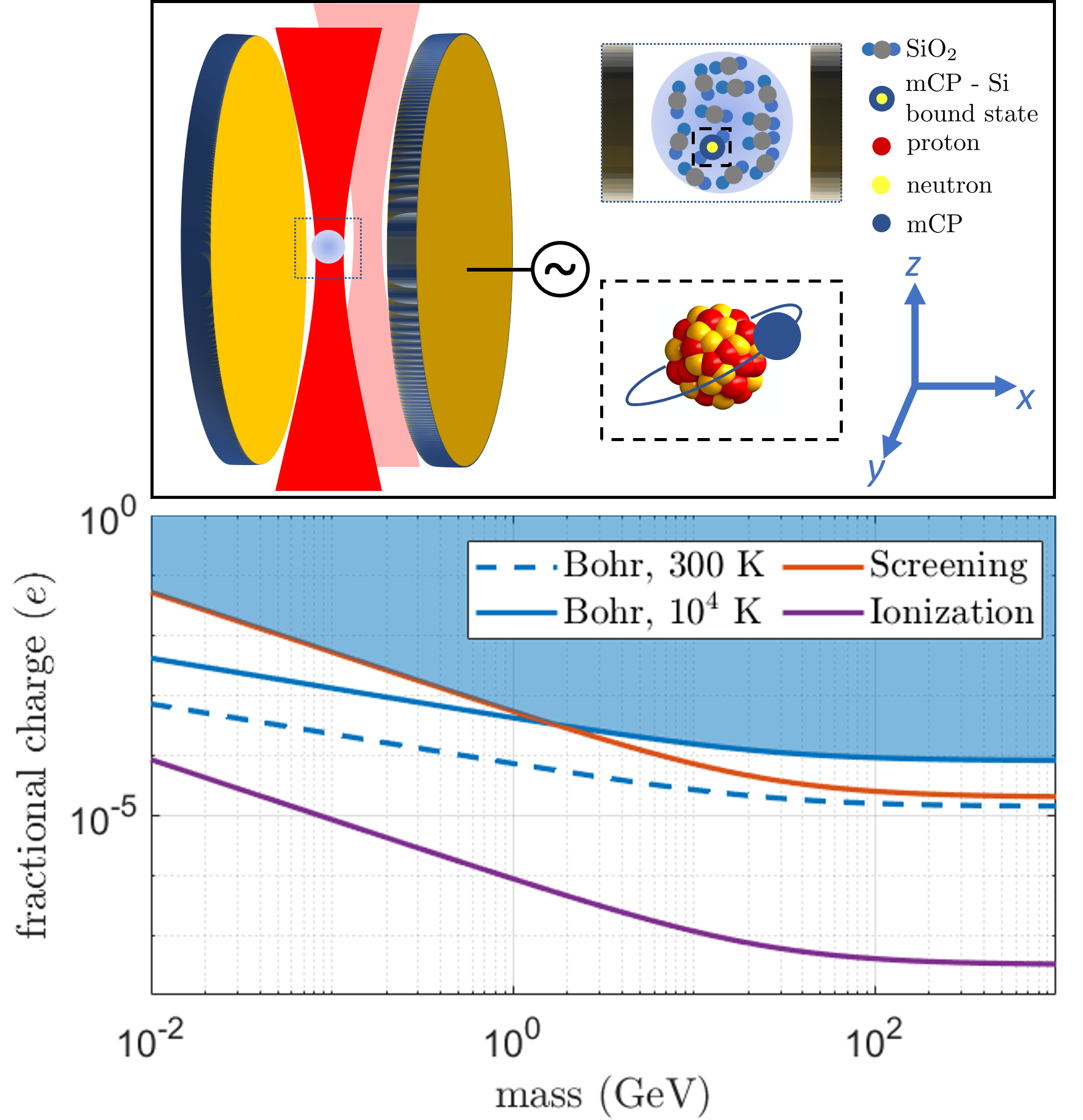}
    \put(220,150){\large \textbf{(a)}}
    \put(220,45){\large \textbf{(b)}}
    \end{overpic}
    \caption[]{(a) SiO$_{2}$ spheres are optically levitated in high vacuum between a pair of parallel electrodes to search for a violation of charge neutrality by, \eg, a mCP electrostatically bound to one of the Si nuclei in the sphere. A second beam (light red) parallel to the trapping beam is used for vibrational background subtraction (see text). (b) Allowed parameter-space for creation of Si-mCP bound-states, for two example minimum binding energies (blue, solid for $E_B/k_B = 10^4$~K, dashed for $E_B/k_B = 300$~K) and  for the requirement that atomic K-shell electron screening is insignificant (red). The shaded region corresponds to the parameter space in which both conditions are simultaneously met. The region above the purple solid line reflects parameter space in which the mCPs are not ionized from the sphere at the highest electric field magnitudes used in the experiment, which is sub-dominant to the binding energy requirements above over the full mass range considered.}
  \label{fig:fig1}
\end{figure}

The approximate region over which stable bound states can be formed was estimated assuming a negatively charged mCP and positively charge Si or O nucleus in the sphere can for a Bohr-like ``atom''~\cite{pospelov2020earthbound} [Fig.~\ref{fig:fig1}~(a), inset]. Stability of such a bound state requires that the binding energy is much larger than the thermal energy (or any other athermal excitation present) in the environment. For a given minimum binding energy, $E_B$, stable bound states can form for $\varepsilon \geq \sqrt{\frac{2E_B}{Z^2\mu \alpha^2}}$. Here $Z$ is the atomic number of the Si or O nucleus and $\mu=mM/(m+M)$ is the reduced mass of the nucleus (mass $M$) and the mCP (mass $m$). The curves corresponding to this limit are shown in Fig~\ref{fig:fig1}~(b) in blue solid and dashed lines for $10^4$~K and 300~K respectively. At lower masses the radius of the bound state can become significant compared to the Bohr radius of the host atom, and  screening effects from the atomic electrons must be considered. As a conservative requirement, we also demand that the lowest orbit of such nucleus-mCP bound states is inside the atomic K-shell~\cite{pospelov2020earthbound}. Practically this means that $\frac{a_0}{Z\varepsilon}\frac{m_e}{\mu}<\frac{a_0}{Z}$, or$\epsilon<\frac{m_e}{\mu}$, where $a_0$ is the atom's Bohr radius and $m_e$ is the electron mass. The curve corresponding to this limit is shown in red. 


In the experiment [Fig.~\ref{fig:fig1} (a)], a vertically oriented 1064~nm laser beam is used to trap 10, 15 and 20~$\mu$m-diameter SiO$_2$ microspheres~\footnote{\url{https://www.microspheres-nanospheres.com/}}. Upon reducing the pressure to $\sim10^{-7}$~mbar, the sphere is neutralized, leaving it with precisely the same number of electrons and protons~\cite{Moore2014,Frimmer2017,Conangla:2018nnn,Monteiro2020}. The sphere is positioned between a parallel pair of electrodes, 25.4~mm in diameter and $2.04\pm0.02$~mm apart, for which the relative tilt is measured to be $\lesssim 1$ mrad. One of the electrodes is connected to a high voltage source ($\pm$10~kV), while the other can be separately biased with low voltage.

\begin{figure} 
  \centering
    \begin{overpic}
      [width=\linewidth]{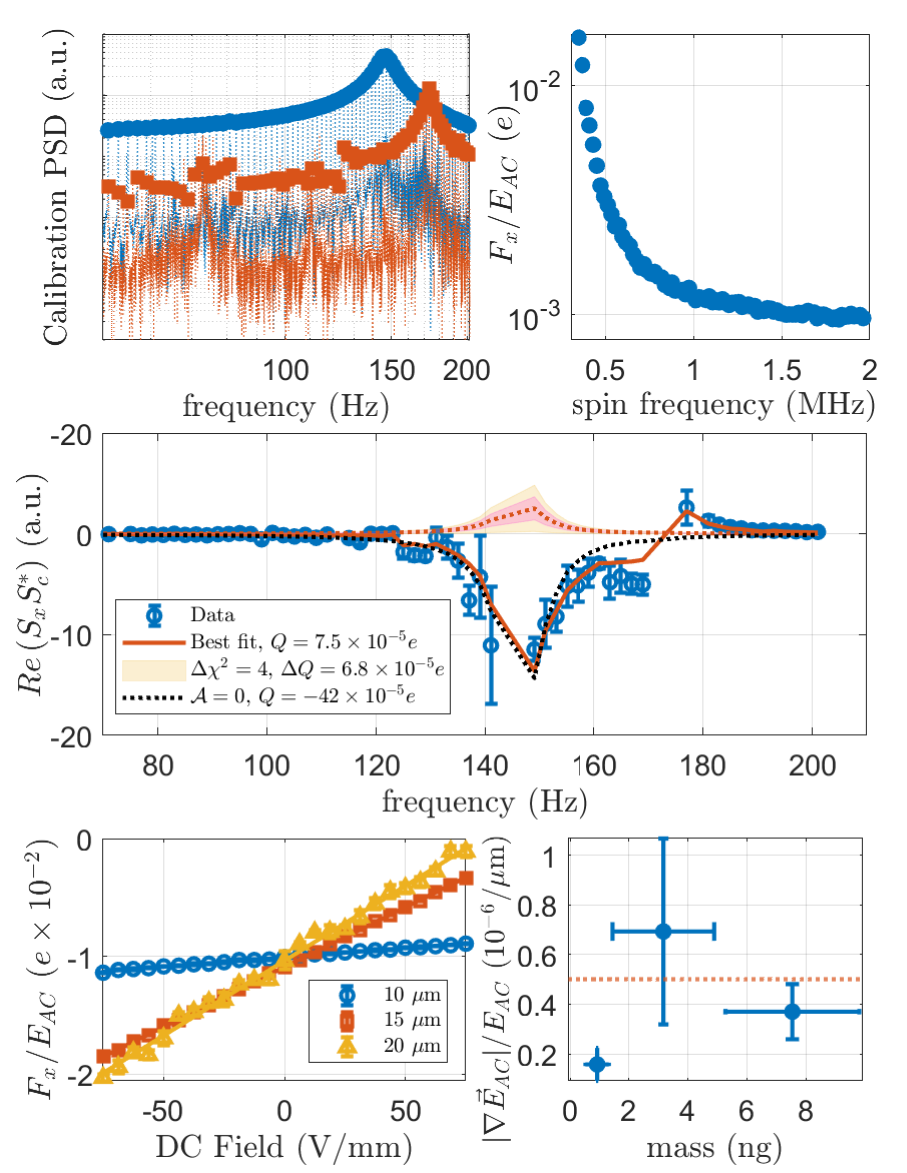}
      \put(30,295){\large \textbf{(a)}}
      \put(215,295){\large \textbf{(b)}}
      \put(30,185){\large \textbf{(c)}}
      \put(30,75){\large \textbf{(d)}}
      \put(215,75){\large \textbf{(e)}}
    \end{overpic}
    \caption[]{Backgrounds. (a) Power spectral density (PSD) of a sphere's motion with net charge of 4$e$, driven by a 100~V/mm frequency comb at odd frequencies between 51 and 201~Hz. Blue markers highlight the drive frequencies. This spectral response is used to subtract the vibrational background (red, squares). (b) Reduction in torque related backgrounds as the rotation rate approaches $\Omega_s\gtrsim1$~MHz. (c) Real part of Eq.~\ref{eq:Vib}. A simultaneous fit is done to both the real and imaginary part, to constrain the leakage of vibrations into the motion that is in-phase with the drive. The value for the excess charge of this sphere is corrected from $Q = -42\times10^{-5}~e$ (black dotted line), to $7.5\pm6.8\times10^{-5}~e$ at 95\% CL (red dotted lines) after profiling over the amplitude of the vibrational background. The red solid line is a fit to the model. (d) Inferred charge for example spheres of each size as a small DC field is varied, from which the gradient at the position of the sphere can be determined. (e) Magnitude of the inferred gradients and comparison to COMSOL simulations (dotted, for $x_j=y,z$).}
  \label{fig:fig2}
\end{figure}
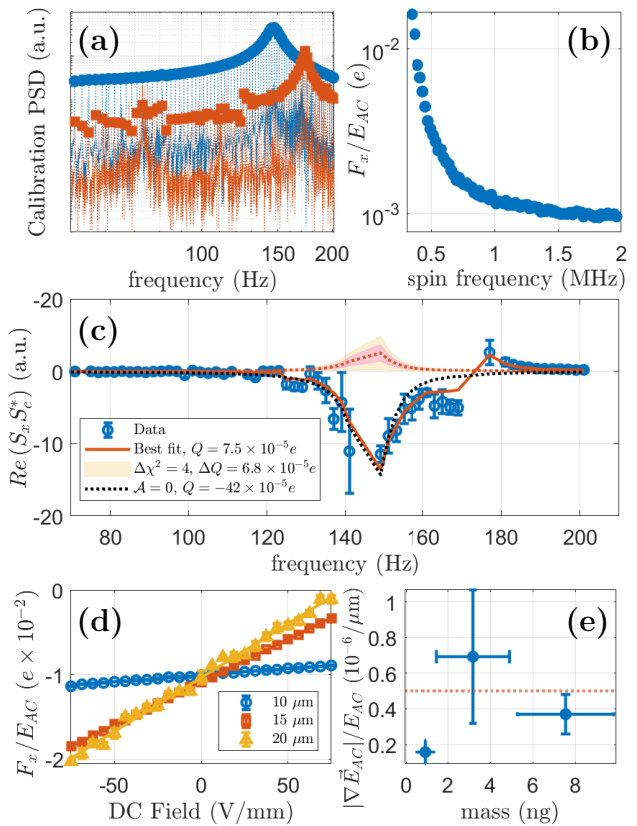

Each sphere's response is calibrated by setting its net charge to $Q = N e$ (for known $N$ between 1--5), and its motion is recorded as it is driven with a frequency comb comprised of equal amplitude tones at odd frequencies between 51 to 201~Hz. Since forces on the induced dipole and most vibrations occur at twice the drive frequency, odd frequencies avoid mixing of backgrounds into the signal. The calibrated response is used as a template representing the motion of a unity-charged sphere under the effect of the drive [Fig.~\ref{fig:fig2} (a), blue]. The sphere is then discharged and the field amplitude increased to typical values of 5~kV/mm, well below the expected threshold for ionization of the mCP which is $\sim2$ orders of magnitude higher [Fig.~\ref{fig:fig1} (b), purple]. The residual motion under the influence of the strong field is correlated against the calibration template in order to the obtain an effective residual charge $\varepsilon$. The presence of a mCP bound to a Si or O nucleus in the sphere [Fig.~\ref{fig:fig1} (a), insets] would induce a response identical to the calibration template, up to rescaling by its charge. In total, \NTen~10-$\mu$m spheres, \NFifteen~15-$\mu$m spheres and \NTwenty~20-$\mu$m spheres were examined, with a total mass of $(76\pm 7)$~ng.


The primary known backgrounds arise from mechanical vibrations that can induce either a real or apparent motion of the sphere that is correlated with the applied voltage, and field-induced torques and forces acting on permanent or induced multipole moments within the sphere. In particular, permanent electric dipole moments coupling to residual electric field gradients are the source of the dominant backgrounds.

Vibrations are generated by the high-voltage drive. Though these mostly occur at the sum and difference of the frequencies of the driving field, they can still, through nonlinearities in the system or a piezoelectric effect in insulating ceramics, leak into the drive frequency and excite the sphere, imitating a charge. A second, weaker 1064~nm beam traverses an optical path similar to that of the trapping beam inside the vacuum chamber and its motion, coupled only to the vibrational background, is recorded [Fig.~\ref{fig:fig2} (a), red] and subtracted from the sphere response after accounting for the corresponding transfer function. This auxiliary beam, depicted in Fig~\ref{fig:fig1}, is parallel to the trapping beam and separated from it by $\sim1$~mm in the trapping plane. Deconvolving the effect of vibrations from the signal is more effective as the resonance of the sphere becomes narrower and more separated from any prominent vibrational peaks. In our system, 15~$\mu$m spheres have a natural resonance close to the vibrational peak at $\sim170$~Hz, and proportional feedback gain is used to reduce their resonance frequency. More formally, the response of the neutralized sphere can be written as $S_x = \mathcal{A}HS_v+QS_c$, where $\mathcal{A}$ is a scaling constant related to the magnitude of the vibrational noise, $H$ is the response function of a driven, damped harmonic oscillator, $Q$ is the a possible residual charge of the sphere and $S_c$ is the spectrum of the unity-charge calibration measurement. $S_v$, the spectrum of the vibrations, is measured using the auxiliary beam described above. A typical vibrational power spectral density (PSD) is shown in red squares in Fig.~\ref{fig:fig2} (a), where it is compared to the PSD of a calibration measurement showing the response of a charged sphere to the same drive, whose spectrum is given by $S_d$. A relation can then be obtained between the spectral response of the sphere, the calibration spectrum and the vibrations:
\begin{align}
\begin{split}\label{eq:Vib}
  \operatorname{Im}(S_xS_c^*) &= \mathcal{A}\vert S_c\vert^2\operatorname{Im}(S_v/S_d)\\ 
  \operatorname{Re}(S_xS_c^*) &= \left[\mathcal{A}\operatorname{Re}(S_v/S_d) + Q\right]\vert S_c\vert^2
\end{split}
\end{align}
The imaginary part of Eq.~\ref{eq:Vib} does not carry any information regarding the charge (i.e., it is out of phase with the charge response, but can be in phase with a portion of the vibration response). In the limit of no vibrations ($\mathcal{A}=0$) the equation for the real part is a simple scaling of the sphere transfer function with charge. Charge is extracted by fitting the real part of Eq.~\ref{eq:Vib} for $Q$, while profiling over a range of possible values for $\mathcal{A}$. For each of these values a $\chi^2$ goodness-of-fit is calculated and the 95\% confidence interval obtained is the reported result. There are three error components to the data $S_xS_c^*$: a statistical error obtained from taking multiple averages, a systematic multiplicative error derived from the discrepancy between the model and the data (derived from the out-of-phase component only) and a small additive systematic of $0.2\times10^{-6}$ in the units of Fig.~\ref{fig:fig2} (c) that accounts for additional deviations seen between the model and out-of-phase data. Fig.~\ref{fig:fig2} (c) shows the data and fits to the real part of Eq.~\ref{eq:Vib} for a 15~$\mu$m sphere, leading to subtraction of motion resulting from vibrational backgrounds by a factor of $\sim5.5$ for this sphere. Typical fits indicate that roughly 50--80\% of the original signal arises from vibrations for the 15~$\mu$m spheres, where the resonance lies near the vibrational peak seen in the co-aligned beam. For the 10 and 20~$\mu$m diameter spheres, the vibration correction is typically negligible.

Forces and torques on the permanent electric dipole moment of the sphere, $\vec{p}_0$, lead to additional backgrounds correlated with the driving field. Torques of the form $\vec{p}_0\times\vec{E}$ can induce angular motion of the sphere which can appear as an apparent center-of-mass motion due to slight asphericity. Torque-induced backgrounds are mitigated by optically spinning the spheres using circularly polarized light at rotational frequencies $\Omega_s\gg1$~MHz~\cite{Arita2013NatCo...4.2374A,Monteiro2018,Ahn2018,Reimann2018}, increasing angular momentum and resistance to precessive tilts. Fig~\ref{fig:fig2} (b) shows a measurement of the inferred ``charge" of a 15~$\mu$m sphere as $\Omega_s$ increases. For $\Omega_s\gtrsim 1$~MHz torques becomes sub-dominant to other backgrounds. The data used here are acquired at $\Omega_s\gg 1$~MHz, where terminal rotation speeds are limited by background gas damping~\cite{Monteiro2018}.

In addition to torques, residual gradients in the applied electric field, $\vec{E}_{AC}$ can lead to a net force, $\vec{F} = (\vec{p}_0\cdot\na) \vec{E}_{AC}$. Gradients are reduced by using electrodes with a diameter much larger than their spacing to eliminate fringing fields and ensuring they are parallel to $0\pm1$~mrad. Finite element simulations of the electric field are performed to obtain the maximal values of $\nabla\vec{E}_{AC}/E_{AC}$ for such a tilt. The gradients at the sphere location are measured by adding and scanning a small DC field, and recording the resultant linear charge-like signal $F_x/E_{AC}$ [Fig~\ref{fig:fig2} (c)]. The slope of this curve is $S = 2\alpha_{xj}\pd{E_x}{x_j}$, where $\alpha_{xx}\approx 4\pi\epsilon_0 R^3\left(\frac{\epsilon-1}{\epsilon+2}\right)$ is the polarizability for an isotropic sphere with dielectric constant $\epsilon$~\cite{Jackson1999}. Due to the small asphericity and amorphous nature of the SiO$_2$ spheres, off-diagonal components of the polarization tensor are expected to be small, \ie $\alpha_{xj} \ll \alpha_{xx}$ for $j\neq x$. The sphere radius, $R$, is typically known to within $\sim10\%$~\cite{Monteiro2017}. Fig~\ref{fig:fig2} (d) shows the measured gradients, which are in agreement with simulations and indicate that the $(\pd{E_x}{x})/E_x\sim10^{-9}$~$\mu$m$^{-1}$ gradient is smaller than those in the $y$ and $z$ directions (red dotted line). Due to lack of knowledge of the precise magnitude and orientation of $\vec{p}_0$, it is not possible to rule out these gradients as the dominant contribution to the sphere response. Previously suggested methods for measuring permanent dipole moments of smaller spheres using a 3D electrode configuration~\cite{Rider:2016xaq,PhysRevA.99.041802_electric_rotation} and future work to measure the direction of $\Omega_s$, which can be affected by the electric field, would enable further reduction of this background. 

Measurements without a sphere in the trap indicate that electrical pickup, scattered light from chamber surfaces, and other technical backgrounds not related to the sphere are negligible. Piezoelectric stretching of the sphere itself in the presence of the applied field could lead to real or apparent forces correlated with the drive. The piezoelectric coefficient required to explain our result is on the order of pm/V, which is comparable to that of crystalline materials such as quartz~\cite{Genevrier2013}, but much higher than expected for amorphous silica. Future searches may be able to identify such effects, \eg by monitoring the frequency of internal optical resonances~\cite{2010PhRvL.105g3002B_barker_wgm}. The magnetic moment required to fully account for an apparent $10^{-4}~e$ charge in a 50~Hz, 10~kV/mm electric field is of the order of $10^{15}~\mu_\text{B}$, where $\mu_\text{B}$ is the Bohr magneton. This is much greater than expected for trace levels of magnetic contaminants in the SiO$_2$. Stray DC electric fields can induce a dipole moment [similar to Fig~\ref{fig:fig2} (c)]. The potential difference between the electrodes required to explain the obtained signal is $\sim17$~V, much larger than the expected $\lesssim 50$~mV from contact potentials or patch potentials on the electrodes~\cite{Garrett:2015bde,Rider:2016xaq}. Higher order multipole moments in the charge distribution are not expected to be significant. Simulations indicate the quadrupole moment required in order to explain our signal is $\sim10^{13}~e\,\mu$m$^2$, which is unphysically large. 

Limits on the average abundance of mCPs bound in the spheres are calculated from the observed data using a profile likelihood based test statistic. The likelihood assumes a Poisson distribution for the expected number of mCPs per sphere with a mean $\mu = n N_\text{nuc}$ where $n$ is the mean abundance (per nucleon) and $N_\text{nuc}$ is the number of nucleons, normalized to the number for a given size sphere for each observed data point.

Conservative constraints are set in the region in which each experiment is sensitive to a single mCP at a given fractional charge. Given the background limited nature of these results (as well as all previous searches for mCPs in matter), sensitivity to charges much smaller than the background-limited sensitivity may require more careful interpretation. However, the same limit setting procedure can also be generalized to fractional charges at which multiple mCPs of a given charge are required to produce an observable signal. For multiple mCPs, two limiting cases are possible: 1) all mCPs bound in matter may carry charges of the same sign (\eg if only bound states between a negative mCP and positively charged nucleus are possible); or 2) an equal number of mCPs of either sign are bound in matter on average (\eg if bound states with mCPs bound either to nuclei or electrons are equally likely). For case 2), while the average charge of a sphere is expected to be zero, fluctuations in the number of positively and negatively charged mCPs (which follow a Skellam distribution) can lead to a non-zero net charge for a given sphere. In reality, the average abundance of different signs of mCPs could also lie between these limiting cases. 

To calculate the likelihood for a given observed sphere charge, the likelihoods for the number of mCPs of each charge are first determined independently (for either assumption 1 or 2 above). The likelihood as a function of the total charge of the sphere is then determined, after accounting for the polarities of the charges. An additional term is added to the likelihood to allow for the presence of a background resulting from forces on the permanent dipole. For each sphere, the dipole contribution is implemented as an additive force constrained by a Gaussian term in the likelihood with zero mean and error equal to the mean response of the sphere (under the assumption that this mean response could correspond either to a dipole force or an actual net charge). The total negative log likelihood (NLL) summed over the corresponding NLL for each sphere is calculated, and the profile likelihood is calculated by minimizing the NLL over the magnitude of the additive background at each mCP abundance. 95\% confidence level limits are determined from the profile likelihood following Wilks' theorem, \ie that $2\times$NLL is distributed following a $\chi^2$ distribution with 1 degree-of-freedom.

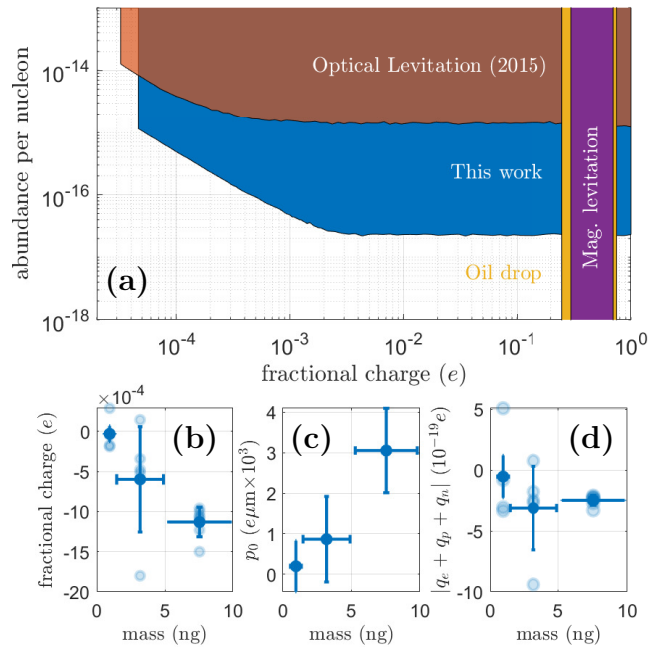
\begin{figure} 
  \centering
    \begin{overpic}
      [width=\linewidth]{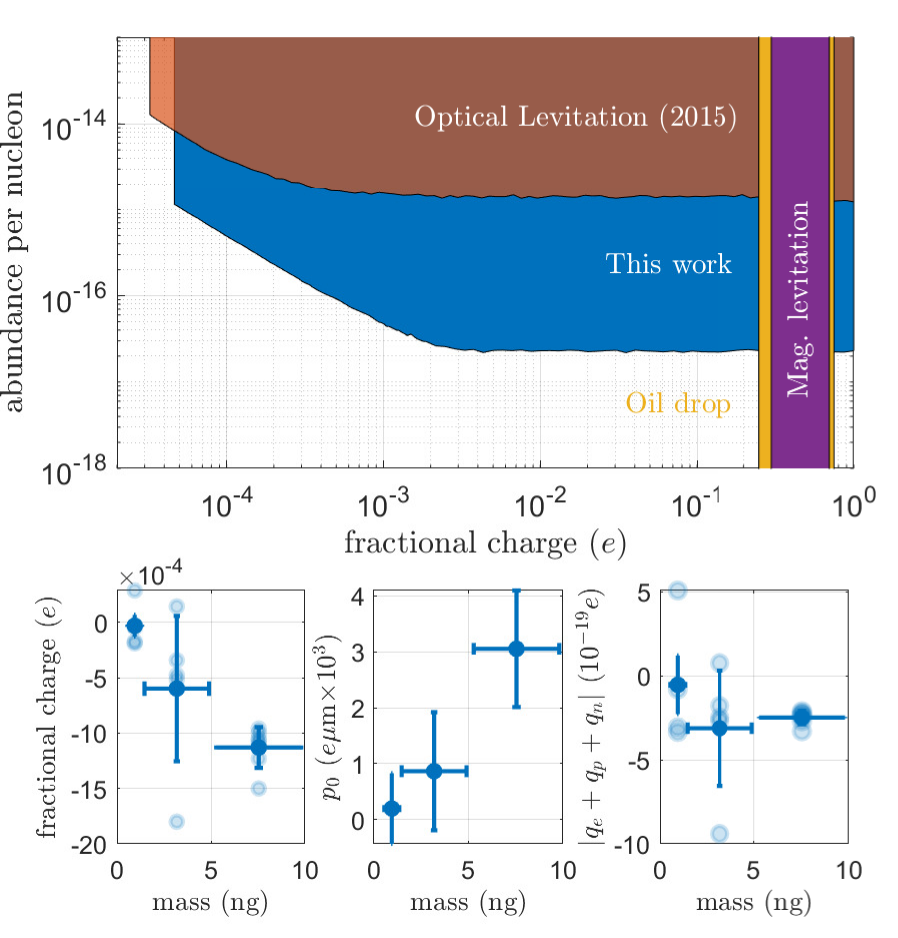}
      \put(35,140){\large \textbf{(a)}}
      \put(60,80){\large \textbf{(b)}}
      \put(105,80){\large \textbf{(c)}}
      \put(210,80){\large \textbf{(d)}}
    \end{overpic}
    \caption[]{(a) 95\% CL limits on abundance in the regions where each experiment is sensitive to a single mCP at a given fractional charge. Results are compared to previous searches in matter~\cite{Marinelli1982,Kim2007,Moore2014}. A $\varepsilon$ smaller than the leftmost edge of the excluded region would require multiple mCPs per sphere. (b) Individual sphere data (light color), along with its mean and standard deviation (bold color). (c) Estimation of the magnitude of the permanent dipole needed to explain the mean signal of (b), using the gradients of Fig~\ref{fig:fig2} (c-d). (d) Limit on the sum of the charges of the proton, electron and neutron at the level of $\sim3 \times 10^{-19}~e$.}
  \label{fig:fig3}
\end{figure}

Fig.~\ref{fig:fig3} (a) shows these limits, compared to previous results obtained using magnetically levitated balls~\cite{Marinelli1982}, Millikan-type oil drops~\cite{Kim2007} and optically levitated 5~$\mu$m-diameter microspheres~\cite{Moore2014}. Solid regions in the plot represent the excluded area for which the experiments are sensitive to a single mCP per sphere. The left edge of the excluded area is set by the sphere with the smallest limit on the measured charge. Below this value, multiple mCPs would be required to produce an observable signature above backgrounds.

A vast majority of spheres [Fig.~\ref{fig:fig3}~(b)] show a response that is consistent with the same sign of charge, and with the mean total charge scaling linearly with mass. Both observations are consistent with mCPs, but could also be potentially explained by a permanent dipole moment held in fixed orientation to the field that scales roughly linearly with the mass of the spheres. In Fig.~\ref{fig:fig3}~(c) we calculate the magnitude of the permanent dipole moment needed to explain the observed results, assuming the measured gradients of Fig~\ref{fig:fig2} (c-d). These values are comparable to recent measurements of the permanent dipole of smaller SiO$_2$ microspheres~\cite{Blakemore2020} assuming a linear scaling in mass. Alternatively, these results could be interpreted as a net charge resulting from a deviation in the sum of the proton, electron, and neutron charges from zero, $Q_m = \vert q_p + q_e + q_n\vert$. This hypothesis has been bound by a value of $\sim10^{-21}~e$~\cite{Dylla1973,Baumann1988,Bressi2011} in previous experiments. Our results, shown in Fig~\ref{fig:fig3} (d), are background-limited to $Q_m \lesssim10^{-19}~e$. The measurement sensitivity from these techniques in the absence of backgrounds is $\sim10^{-21}~e/\sqrt{\mathrm{Hz}}$~\cite{Monteiro2020}, similar to the current best limits. Further understanding of backgrounds would enable enhanced sensitivity to this fundamental property of matter.

\begin{figure} 
  \centering
    \begin{overpic}
      [width=\linewidth]{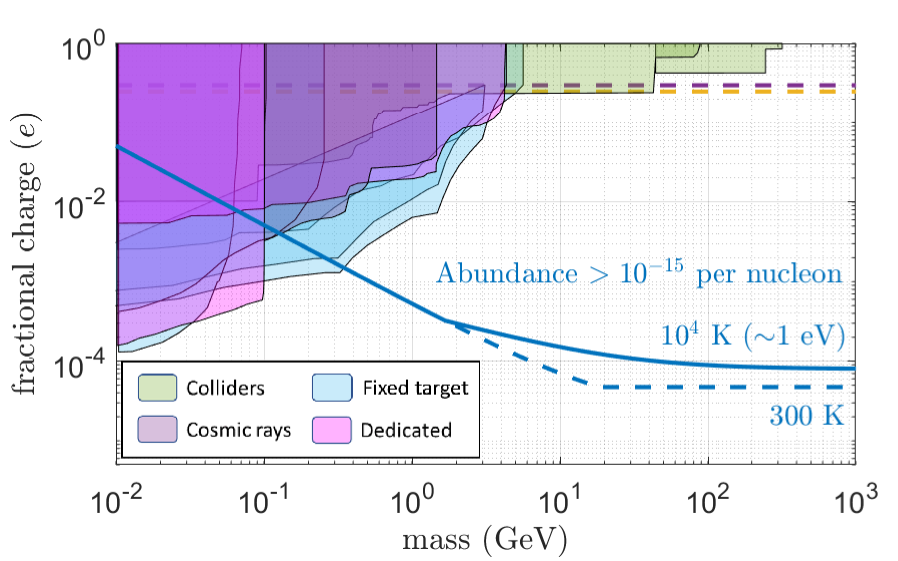}
    \end{overpic}
    \caption[]{For an abundance $\gtrsim 10^{-15}$ per nucleon, limits are placed on the charge vs. the mass of mCPs for a binding energy, $E_B \geq k_B T$ at the temperatures shown. The result of the previous search in levitated spheres~\cite{Moore2014} would yield similar ($10^4$~K) or slightly better (300~K) limits at higher abundances. These constraints are compared to existing limits on mCPs~\cite{Davidson2000LEP,Jaeckel2013LHC,CMS:2012xi,Chatrchyan:2013oca,Acciarri2020ArgoNeuT,Magill:2018tbb,Prinz1998SLAC,Ball2020MilliquanDemo,Marocco2021}, as well as previous searches in matter~\cite{Marinelli1982,Kim2007} (dashed, colors corresponding to Fig.~\ref{fig:fig3}). Translations of limits from neutrino~\cite{Alvis:2018yte,Singh:2018von} and DM~\cite{alkhatib2020constraints,Agnese:2014vxh} experiments on the underground flux of mCPs are not shown since they are not expected to be sensitive to relic DM mCPs~\cite{alkhatib2020constraints}, and analyses of the expected cosmic-ray flux provide weaker constraints than the direct searches shown~\cite{Plestid:2020kdm}. For a given $E_B$, constraints are mCP-mass dependent since lower mass mCPs are more easily ionized.}
  \label{fig:fig4}
\end{figure}

For $\vert\varepsilon\vert > 4.7\times10^{-5}~e$ the current limits improve sensitivity to single mCPs bound in matter by 1--2 orders-of-magnitude, reaching abundance of $\lesssim2\times 10^{-17}$ mCPs per nucleon for fractional charges $\gtrsim10^{-3}~e$. Such limits can be translated to constraints on the fractional charge of mCPs versus their mass (Fig.~\ref{fig:fig4}), assuming an abundance in surface terrestrial matter $\gtrsim 10^{-15}$ per nucleon and under the requirement that a mCP/nucleus bound state has sufficient binding energy to remain stable during the production of the sphere and the measurement of its charge, taking into account possible screening by atomic electrons.  

If such particles are present in matter at this abundance, and assuming that the Bohr binding energy $E_B$ is above a given ionization threshold of $300$~K and $10^4$~K ($\approx 1$~eV, typical for atoms which are stable even in circumstances where athermal excitations are common), the results shown in Fig.~\ref{fig:fig4} explore a substantial fraction of parameter space beyond the reach of current and future accelerator and underground experiments. Such abundances are possible \eg from capture of relic DM in terrestrial materials, where the per-nucleon abundance sensitivity of $\gtrsim 10^{-17}$ would probe sub-fractions of relic particles as small as $f_Q \gtrsim 10^{-7}$, if the silica used here (expected to be sourced from surface, or near-surface materials) contains mCPs at or above the estimated mean terrestrial abundance for $\sim$GeV mass particles~\cite{pospelov2020earthbound}. Limits on mCPs from cosmology have also been studied~\cite{Dubovsky2004WMAP,Dolgov:2013una,Boddy:2018wzy}, although they typically do not constrain relic DM mCPs if they make up a $\lesssim 0.1$\% sub-component of DM~\cite{Dolgov:2013una,Boddy:2018wzy}, while the results presented here can be sensitive to substantially smaller sub-components. 


In summary, we have presented a search for tiny electric charges, as small as $10^{-4}~e$, in a total mass of $\sim76$~ng. Such charges could arise from mCPs bound to atomic matter or a violation of the neutrality of matter. Our results improve current state-of-the-art limits on the abundance of mCPs bound to matter by approximately two orders-of-magnitude. Assuming an abundance of $\gtrsim10^{-17}-\gtrsim10^{-15}$ mCPs per nucleon, the search presented here probes a substantial gap in parameter space beyond the reach of existing and planned searches~\cite{Magill:2018tbb}. Limits are set on the violation of the neutrality of matter at the $10^{-19}~e$ level. A number of background sources to such measurements are investigated in detail, and future work to improve understanding of backgrounds may enable searches using these techniques to provide further advances in searches for fractional charges in matter.

\begin{acknowledgments}
The authors would like to thank Yotam Soreq (Technion), 
Harikrishnan Ramani and the Gratta group (Stanford) for valuable discussions. This work is supported, in part, by NSF Grant PHY-1653232, the Heising-Simons Foundation, the Alfred P. Sloan Foundation, and ONR Grant N00014-18-1-2409.
\end{acknowledgments}

\bibliographystyle{apsrev4-1}
\bibliography{Millicharge}

\begin{thebibliography}{74}%
\makeatletter
\providecommand \@ifxundefined [1]{%
 \@ifx{#1\undefined}
}%
\providecommand \@ifnum [1]{%
 \ifnum #1\expandafter \@firstoftwo
 \else \expandafter \@secondoftwo
 \fi
}%
\providecommand \@ifx [1]{%
 \ifx #1\expandafter \@firstoftwo
 \else \expandafter \@secondoftwo
 \fi
}%
\providecommand \natexlab [1]{#1}%
\providecommand \enquote  [1]{``#1''}%
\providecommand \bibnamefont  [1]{#1}%
\providecommand \bibfnamefont [1]{#1}%
\providecommand \citenamefont [1]{#1}%
\providecommand \href@noop [0]{\@secondoftwo}%
\providecommand \href [0]{\begingroup \@sanitize@url \@href}%
\providecommand \@href[1]{\@@startlink{#1}\@@href}%
\providecommand \@@href[1]{\endgroup#1\@@endlink}%
\providecommand \@sanitize@url [0]{\catcode `\\12\catcode `\$12\catcode
  `\&12\catcode `\#12\catcode `\^12\catcode `\_12\catcode `\%12\relax}%
\providecommand \@@startlink[1]{}%
\providecommand \@@endlink[0]{}%
\providecommand \url  [0]{\begingroup\@sanitize@url \@url }%
\providecommand \@url [1]{\endgroup\@href {#1}{\urlprefix }}%
\providecommand \urlprefix  [0]{URL }%
\providecommand \Eprint [0]{\href }%
\providecommand \doibase [0]{http://dx.doi.org/}%
\providecommand \selectlanguage [0]{\@gobble}%
\providecommand \bibinfo  [0]{\@secondoftwo}%
\providecommand \bibfield  [0]{\@secondoftwo}%
\providecommand \translation [1]{[#1]}%
\providecommand \BibitemOpen [0]{}%
\providecommand \bibitemStop [0]{}%
\providecommand \bibitemNoStop [0]{.\EOS\space}%
\providecommand \EOS [0]{\spacefactor3000\relax}%
\providecommand \BibitemShut  [1]{\csname bibitem#1\endcsname}%
\let\auto@bib@innerbib\@empty
\bibitem [{\citenamefont {Millikan.}(1913)}]{PhysRev.2.109_millikan}%
  \BibitemOpen
  \bibfield  {author} {\bibinfo {author} {\bibfnamefont {R.~A.}\ \bibnamefont
  {Millikan.}},\ }\href {\doibase 10.1103/PhysRev.2.109} {\bibfield  {journal}
  {\bibinfo  {journal} {Phys. Rev.}\ }\textbf {\bibinfo {volume} {2}},\
  \bibinfo {pages} {109} (\bibinfo {year} {1913})}\BibitemShut {NoStop}%
\bibitem [{\citenamefont {Unnikrishnan}\ and\ \citenamefont
  {Gillies}(2004)}]{Unnikrishnan_2004}%
  \BibitemOpen
  \bibfield  {author} {\bibinfo {author} {\bibfnamefont {C.~S.}\ \bibnamefont
  {Unnikrishnan}}\ and\ \bibinfo {author} {\bibfnamefont {G.~T.}\ \bibnamefont
  {Gillies}},\ }\href {\doibase 10.1088/0026-1394/41/5/s03} {\bibfield
  {journal} {\bibinfo  {journal} {Metrologia}\ }\textbf {\bibinfo {volume}
  {41}},\ \bibinfo {pages} {S125} (\bibinfo {year} {2004})}\BibitemShut
  {NoStop}%
\bibitem [{\citenamefont {Dobroliubov}\ and\ \citenamefont
  {Ignatiev}(1990)}]{PhysRevLett.65.679_mcp}%
  \BibitemOpen
  \bibfield  {author} {\bibinfo {author} {\bibfnamefont {M.~I.}\ \bibnamefont
  {Dobroliubov}}\ and\ \bibinfo {author} {\bibfnamefont {A.~Y.}\ \bibnamefont
  {Ignatiev}},\ }\href {\doibase 10.1103/PhysRevLett.65.679} {\bibfield
  {journal} {\bibinfo  {journal} {Phys. Rev. Lett.}\ }\textbf {\bibinfo
  {volume} {65}},\ \bibinfo {pages} {679} (\bibinfo {year} {1990})}\BibitemShut
  {NoStop}%
\bibitem [{\citenamefont {Jaeckel}\ and\ \citenamefont
  {Ringwald}(2010)}]{Jaeckel:2010ni_review}%
  \BibitemOpen
  \bibfield  {author} {\bibinfo {author} {\bibfnamefont {J.}~\bibnamefont
  {Jaeckel}}\ and\ \bibinfo {author} {\bibfnamefont {A.}~\bibnamefont
  {Ringwald}},\ }\href {\doibase 10.1146/annurev.nucl.012809.104433} {\bibfield
   {journal} {\bibinfo  {journal} {Ann. Rev. Nucl. Part. Sci.}\ }\textbf
  {\bibinfo {volume} {60}},\ \bibinfo {pages} {405} (\bibinfo {year} {2010})},\
  \Eprint {http://arxiv.org/abs/1002.0329} {arXiv:1002.0329 [hep-ph]}
  \BibitemShut {NoStop}%
\bibitem [{\citenamefont {Holdom}(1986)}]{Holdom1986}%
  \BibitemOpen
  \bibfield  {author} {\bibinfo {author} {\bibfnamefont {B.}~\bibnamefont
  {Holdom}},\ }\href {\doibase https://doi.org/10.1016/0370-2693(86)91377-8}
  {\bibfield  {journal} {\bibinfo  {journal} {Phys. Lett. B}\ }\textbf
  {\bibinfo {volume} {166}},\ \bibinfo {pages} {196 } (\bibinfo {year}
  {1986})}\BibitemShut {NoStop}%
\bibitem [{\citenamefont {Izaguirre}\ and\ \citenamefont
  {Yavin}(2015)}]{PhysRevD.92.035014_mCP_hypercharge_window}%
  \BibitemOpen
  \bibfield  {author} {\bibinfo {author} {\bibfnamefont {E.}~\bibnamefont
  {Izaguirre}}\ and\ \bibinfo {author} {\bibfnamefont {I.}~\bibnamefont
  {Yavin}},\ }\href {\doibase 10.1103/PhysRevD.92.035014} {\bibfield  {journal}
  {\bibinfo  {journal} {Phys. Rev. D}\ }\textbf {\bibinfo {volume} {92}},\
  \bibinfo {pages} {035014} (\bibinfo {year} {2015})}\BibitemShut {NoStop}%
\bibitem [{\citenamefont {Brahm}\ and\ \citenamefont
  {Hall}(1990)}]{PhysRevD.41.1067_old_dark_photon}%
  \BibitemOpen
  \bibfield  {author} {\bibinfo {author} {\bibfnamefont {D.~E.}\ \bibnamefont
  {Brahm}}\ and\ \bibinfo {author} {\bibfnamefont {L.~J.}\ \bibnamefont
  {Hall}},\ }\href {\doibase 10.1103/PhysRevD.41.1067} {\bibfield  {journal}
  {\bibinfo  {journal} {Phys. Rev. D}\ }\textbf {\bibinfo {volume} {41}},\
  \bibinfo {pages} {1067} (\bibinfo {year} {1990})}\BibitemShut {NoStop}%
\bibitem [{\citenamefont {Pospelov}\ \emph {et~al.}(2008)\citenamefont
  {Pospelov}, \citenamefont {Ritz},\ and\ \citenamefont
  {Voloshin}}]{Pospelov:2007mp}%
  \BibitemOpen
  \bibfield  {author} {\bibinfo {author} {\bibfnamefont {M.}~\bibnamefont
  {Pospelov}}, \bibinfo {author} {\bibfnamefont {A.}~\bibnamefont {Ritz}}, \
  and\ \bibinfo {author} {\bibfnamefont {M.~B.}\ \bibnamefont {Voloshin}},\
  }\href {\doibase 10.1016/j.physletb.2008.02.052} {\bibfield  {journal}
  {\bibinfo  {journal} {Phys. Lett. B}\ }\textbf {\bibinfo {volume} {662}},\
  \bibinfo {pages} {53} (\bibinfo {year} {2008})},\ \Eprint
  {http://arxiv.org/abs/0711.4866} {arXiv:0711.4866 [hep-ph]} \BibitemShut
  {NoStop}%
\bibitem [{\citenamefont {Bjorken}\ \emph {et~al.}(2009)\citenamefont
  {Bjorken}, \citenamefont {Essig}, \citenamefont {Schuster},\ and\
  \citenamefont {Toro}}]{PhysRevD.80.075018_accel_dark_forces}%
  \BibitemOpen
  \bibfield  {author} {\bibinfo {author} {\bibfnamefont {J.~D.}\ \bibnamefont
  {Bjorken}}, \bibinfo {author} {\bibfnamefont {R.}~\bibnamefont {Essig}},
  \bibinfo {author} {\bibfnamefont {P.}~\bibnamefont {Schuster}}, \ and\
  \bibinfo {author} {\bibfnamefont {N.}~\bibnamefont {Toro}},\ }\href {\doibase
  10.1103/PhysRevD.80.075018} {\bibfield  {journal} {\bibinfo  {journal} {Phys.
  Rev. D}\ }\textbf {\bibinfo {volume} {80}},\ \bibinfo {pages} {075018}
  (\bibinfo {year} {2009})}\BibitemShut {NoStop}%
\bibitem [{\citenamefont {Izaguirre}\ \emph {et~al.}(2013)\citenamefont
  {Izaguirre}, \citenamefont {Krnjaic}, \citenamefont {Schuster},\ and\
  \citenamefont {Toro}}]{PhysRevD.88.114015_gordan_accel}%
  \BibitemOpen
  \bibfield  {author} {\bibinfo {author} {\bibfnamefont {E.}~\bibnamefont
  {Izaguirre}}, \bibinfo {author} {\bibfnamefont {G.}~\bibnamefont {Krnjaic}},
  \bibinfo {author} {\bibfnamefont {P.}~\bibnamefont {Schuster}}, \ and\
  \bibinfo {author} {\bibfnamefont {N.}~\bibnamefont {Toro}},\ }\href {\doibase
  10.1103/PhysRevD.88.114015} {\bibfield  {journal} {\bibinfo  {journal} {Phys.
  Rev. D}\ }\textbf {\bibinfo {volume} {88}},\ \bibinfo {pages} {114015}
  (\bibinfo {year} {2013})}\BibitemShut {NoStop}%
\bibitem [{\citenamefont {Foot}\ \emph {et~al.}(1993)\citenamefont {Foot},
  \citenamefont {Lew},\ and\ \citenamefont {Volkas}}]{Foot_1993}%
  \BibitemOpen
  \bibfield  {author} {\bibinfo {author} {\bibfnamefont {R.}~\bibnamefont
  {Foot}}, \bibinfo {author} {\bibfnamefont {H.}~\bibnamefont {Lew}}, \ and\
  \bibinfo {author} {\bibfnamefont {R.~R.}\ \bibnamefont {Volkas}},\ }\href
  {\doibase 10.1088/0954-3899/19/3/005} {\bibfield  {journal} {\bibinfo
  {journal} {J. Phys. G: Nucl. and Part. Phys.}\ }\textbf {\bibinfo {volume}
  {19}},\ \bibinfo {pages} {361} (\bibinfo {year} {1993})}\BibitemShut
  {NoStop}%
\bibitem [{\citenamefont {Langacker}(1981)}]{LANGACKER1981185_GUT_review}%
  \BibitemOpen
  \bibfield  {author} {\bibinfo {author} {\bibfnamefont {P.}~\bibnamefont
  {Langacker}},\ }\href {\doibase https://doi.org/10.1016/0370-1573(81)90059-4}
  {\bibfield  {journal} {\bibinfo  {journal} {Phys. Rep.}\ }\textbf {\bibinfo
  {volume} {72}},\ \bibinfo {pages} {185 } (\bibinfo {year}
  {1981})}\BibitemShut {NoStop}%
\bibitem [{\citenamefont {Foot}(1994)}]{PhysRevD.49.3617_foot_anomaly}%
  \BibitemOpen
  \bibfield  {author} {\bibinfo {author} {\bibfnamefont {R.}~\bibnamefont
  {Foot}},\ }\href {\doibase 10.1103/PhysRevD.49.3617} {\bibfield  {journal}
  {\bibinfo  {journal} {Phys. Rev. D}\ }\textbf {\bibinfo {volume} {49}},\
  \bibinfo {pages} {3617} (\bibinfo {year} {1994})}\BibitemShut {NoStop}%
\bibitem [{\citenamefont {{Millen}}\ \emph {et~al.}(2020)\citenamefont
  {{Millen}}, \citenamefont {{Monteiro}}, \citenamefont {{Pettit}},\ and\
  \citenamefont {{Vamivakas}}}]{2020RPPh...83b6401M_millen_review}%
  \BibitemOpen
  \bibfield  {author} {\bibinfo {author} {\bibfnamefont {J.}~\bibnamefont
  {{Millen}}}, \bibinfo {author} {\bibfnamefont {T.~S.}\ \bibnamefont
  {{Monteiro}}}, \bibinfo {author} {\bibfnamefont {R.}~\bibnamefont
  {{Pettit}}}, \ and\ \bibinfo {author} {\bibfnamefont {A.~N.}\ \bibnamefont
  {{Vamivakas}}},\ }\href {\doibase 10.1088/1361-6633/ab6100} {\bibfield
  {journal} {\bibinfo  {journal} {Rep. Prog. Phys.}\ }\textbf {\bibinfo
  {volume} {83}},\ \bibinfo {eid} {026401} (\bibinfo {year}
  {2020})}\BibitemShut {NoStop}%
\bibitem [{\citenamefont {Ahn}\ \emph {et~al.}(2020)\citenamefont {Ahn},
  \citenamefont {Xu}, \citenamefont {Bang}, \citenamefont {Ju}, \citenamefont
  {Gao},\ and\ \citenamefont {Li}}]{Ahn2020}%
  \BibitemOpen
  \bibfield  {author} {\bibinfo {author} {\bibfnamefont {J.}~\bibnamefont
  {Ahn}}, \bibinfo {author} {\bibfnamefont {Z.}~\bibnamefont {Xu}}, \bibinfo
  {author} {\bibfnamefont {J.}~\bibnamefont {Bang}}, \bibinfo {author}
  {\bibfnamefont {P.}~\bibnamefont {Ju}}, \bibinfo {author} {\bibfnamefont
  {X.}~\bibnamefont {Gao}}, \ and\ \bibinfo {author} {\bibfnamefont
  {T.}~\bibnamefont {Li}},\ }\href@noop {} {\bibfield  {journal} {\bibinfo
  {journal} {Nature Nano.}\ }\textbf {\bibinfo {volume} {15}},\ \bibinfo
  {pages} {89} (\bibinfo {year} {2020})}\BibitemShut {NoStop}%
\bibitem [{\citenamefont {Moore}\ and\ \citenamefont
  {Geraci}(2021)}]{Moore:2020awi}%
  \BibitemOpen
  \bibfield  {author} {\bibinfo {author} {\bibfnamefont {D.~C.}\ \bibnamefont
  {Moore}}\ and\ \bibinfo {author} {\bibfnamefont {A.~A.}\ \bibnamefont
  {Geraci}},\ }\href {\doibase 10.1088/2058-9565/abcf8a} {\bibfield  {journal}
  {\bibinfo  {journal} {Quantum Science and Technology}\ }\textbf {\bibinfo
  {volume} {6}},\ \bibinfo {pages} {014008} (\bibinfo {year}
  {2021})}\BibitemShut {NoStop}%
\bibitem [{\citenamefont {Lewandowski}\ \emph {et~al.}(2021)\citenamefont
  {Lewandowski}, \citenamefont {Knowles}, \citenamefont {Etienne},\ and\
  \citenamefont {D'Urso}}]{lewandowski2020high}%
  \BibitemOpen
  \bibfield  {author} {\bibinfo {author} {\bibfnamefont {C.~W.}\ \bibnamefont
  {Lewandowski}}, \bibinfo {author} {\bibfnamefont {T.~D.}\ \bibnamefont
  {Knowles}}, \bibinfo {author} {\bibfnamefont {Z.~B.}\ \bibnamefont
  {Etienne}}, \ and\ \bibinfo {author} {\bibfnamefont {B.}~\bibnamefont
  {D'Urso}},\ }\href {\doibase 10.1103/PhysRevApplied.15.014050} {\bibfield
  {journal} {\bibinfo  {journal} {Phys. Rev. Applied}\ }\textbf {\bibinfo
  {volume} {15}},\ \bibinfo {pages} {014050} (\bibinfo {year}
  {2021})}\BibitemShut {NoStop}%
\bibitem [{\citenamefont {Vinante}\ \emph {et~al.}(2020)\citenamefont
  {Vinante}, \citenamefont {Gasbarri}, \citenamefont {Timberlake},
  \citenamefont {Toro\ifmmode~\check{s}\else \v{s}\fi{}},\ and\ \citenamefont
  {Ulbricht}}]{Vinante2020}%
  \BibitemOpen
  \bibfield  {author} {\bibinfo {author} {\bibfnamefont {A.}~\bibnamefont
  {Vinante}}, \bibinfo {author} {\bibfnamefont {G.}~\bibnamefont {Gasbarri}},
  \bibinfo {author} {\bibfnamefont {C.}~\bibnamefont {Timberlake}}, \bibinfo
  {author} {\bibfnamefont {M.}~\bibnamefont {Toro\ifmmode~\check{s}\else
  \v{s}\fi{}}}, \ and\ \bibinfo {author} {\bibfnamefont {H.}~\bibnamefont
  {Ulbricht}},\ }\href {\doibase 10.1103/PhysRevResearch.2.043229} {\bibfield
  {journal} {\bibinfo  {journal} {Phys. Rev. Research}\ }\textbf {\bibinfo
  {volume} {2}},\ \bibinfo {pages} {043229} (\bibinfo {year}
  {2020})}\BibitemShut {NoStop}%
\bibitem [{\citenamefont {Tebbenjohanns}\ \emph {et~al.}(2019)\citenamefont
  {Tebbenjohanns}, \citenamefont {Frimmer}, \citenamefont {Militaru},
  \citenamefont {Jain},\ and\ \citenamefont {Novotny}}]{Tebbenjohanns2019}%
  \BibitemOpen
  \bibfield  {author} {\bibinfo {author} {\bibfnamefont {F.}~\bibnamefont
  {Tebbenjohanns}}, \bibinfo {author} {\bibfnamefont {M.}~\bibnamefont
  {Frimmer}}, \bibinfo {author} {\bibfnamefont {A.}~\bibnamefont {Militaru}},
  \bibinfo {author} {\bibfnamefont {V.}~\bibnamefont {Jain}}, \ and\ \bibinfo
  {author} {\bibfnamefont {L.}~\bibnamefont {Novotny}},\ }\href {\doibase
  10.1103/PhysRevLett.122.223601} {\bibfield  {journal} {\bibinfo  {journal}
  {Phys. Rev. Lett.}\ }\textbf {\bibinfo {volume} {122}},\ \bibinfo {pages}
  {223601} (\bibinfo {year} {2019})}\BibitemShut {NoStop}%
\bibitem [{\citenamefont {Kamba}\ \emph {et~al.}(2020)\citenamefont {Kamba},
  \citenamefont {Kiuchi}, \citenamefont {Yotsuya},\ and\ \citenamefont
  {Aikawa}}]{Kamba2020}%
  \BibitemOpen
  \bibfield  {author} {\bibinfo {author} {\bibfnamefont {M.}~\bibnamefont
  {Kamba}}, \bibinfo {author} {\bibfnamefont {H.}~\bibnamefont {Kiuchi}},
  \bibinfo {author} {\bibfnamefont {T.}~\bibnamefont {Yotsuya}}, \ and\
  \bibinfo {author} {\bibfnamefont {K.}~\bibnamefont {Aikawa}},\ }\href@noop {}
  {\bibfield  {journal} {\bibinfo  {journal} {arXiv preprint arXiv:2011.12507}\
  } (\bibinfo {year} {2020})}\BibitemShut {NoStop}%
\bibitem [{\citenamefont {Deli{\'c}}\ \emph {et~al.}(2020)\citenamefont
  {Deli{\'c}}, \citenamefont {Reisenbauer}, \citenamefont {Dare}, \citenamefont
  {Grass}, \citenamefont {Vuleti{\'c}}, \citenamefont {Kiesel},\ and\
  \citenamefont {Aspelmeyer}}]{Delic2020}%
  \BibitemOpen
  \bibfield  {author} {\bibinfo {author} {\bibfnamefont {U.}~\bibnamefont
  {Deli{\'c}}}, \bibinfo {author} {\bibfnamefont {M.}~\bibnamefont
  {Reisenbauer}}, \bibinfo {author} {\bibfnamefont {K.}~\bibnamefont {Dare}},
  \bibinfo {author} {\bibfnamefont {D.}~\bibnamefont {Grass}}, \bibinfo
  {author} {\bibfnamefont {V.}~\bibnamefont {Vuleti{\'c}}}, \bibinfo {author}
  {\bibfnamefont {N.}~\bibnamefont {Kiesel}}, \ and\ \bibinfo {author}
  {\bibfnamefont {M.}~\bibnamefont {Aspelmeyer}},\ }\href {\doibase
  10.1126/science.aba3993} {\bibfield  {journal} {\bibinfo  {journal}
  {Science}\ }\textbf {\bibinfo {volume} {367}},\ \bibinfo {pages} {892}
  (\bibinfo {year} {2020})}\BibitemShut {NoStop}%
\bibitem [{\citenamefont {Carney}\ \emph {et~al.}(2021)\citenamefont {Carney},
  \citenamefont {Krnjaic}, \citenamefont {Moore}, \citenamefont {Regal},
  \citenamefont {Afek}, \citenamefont {Bhave}, \citenamefont {Brubaker},
  \citenamefont {Corbitt}, \citenamefont {Cripe}, \citenamefont {Crisosto},
  \citenamefont {Geraci}, \citenamefont {Ghosh}, \citenamefont {Harris},
  \citenamefont {Hook}, \citenamefont {Kolb}, \citenamefont {Kunjummen},
  \citenamefont {Lang}, \citenamefont {Li}, \citenamefont {Lin}, \citenamefont
  {Liu}, \citenamefont {Lykken}, \citenamefont {Magrini}, \citenamefont
  {Manley}, \citenamefont {Matsumoto}, \citenamefont {Monte}, \citenamefont
  {Monteiro}, \citenamefont {Purdy}, \citenamefont {Riedel}, \citenamefont
  {Singh}, \citenamefont {Singh}, \citenamefont {Sinha}, \citenamefont
  {Taylor}, \citenamefont {Qin}, \citenamefont {Wilson},\ and\ \citenamefont
  {Zhao}}]{Carney2020}%
  \BibitemOpen
  \bibfield  {author} {\bibinfo {author} {\bibfnamefont {D.}~\bibnamefont
  {Carney}}, \bibinfo {author} {\bibfnamefont {G.}~\bibnamefont {Krnjaic}},
  \bibinfo {author} {\bibfnamefont {D.~C.}\ \bibnamefont {Moore}}, \bibinfo
  {author} {\bibfnamefont {C.~A.}\ \bibnamefont {Regal}}, \bibinfo {author}
  {\bibfnamefont {G.}~\bibnamefont {Afek}}, \bibinfo {author} {\bibfnamefont
  {S.}~\bibnamefont {Bhave}}, \bibinfo {author} {\bibfnamefont
  {B.}~\bibnamefont {Brubaker}}, \bibinfo {author} {\bibfnamefont
  {T.}~\bibnamefont {Corbitt}}, \bibinfo {author} {\bibfnamefont
  {J.}~\bibnamefont {Cripe}}, \bibinfo {author} {\bibfnamefont
  {N.}~\bibnamefont {Crisosto}}, \bibinfo {author} {\bibfnamefont
  {A.}~\bibnamefont {Geraci}}, \bibinfo {author} {\bibfnamefont
  {S.}~\bibnamefont {Ghosh}}, \bibinfo {author} {\bibfnamefont {J.~G.~E.}\
  \bibnamefont {Harris}}, \bibinfo {author} {\bibfnamefont {A.}~\bibnamefont
  {Hook}}, \bibinfo {author} {\bibfnamefont {E.~W.}\ \bibnamefont {Kolb}},
  \bibinfo {author} {\bibfnamefont {J.}~\bibnamefont {Kunjummen}}, \bibinfo
  {author} {\bibfnamefont {R.~F.}\ \bibnamefont {Lang}}, \bibinfo {author}
  {\bibfnamefont {T.}~\bibnamefont {Li}}, \bibinfo {author} {\bibfnamefont
  {T.}~\bibnamefont {Lin}}, \bibinfo {author} {\bibfnamefont {Z.}~\bibnamefont
  {Liu}}, \bibinfo {author} {\bibfnamefont {J.}~\bibnamefont {Lykken}},
  \bibinfo {author} {\bibfnamefont {L.}~\bibnamefont {Magrini}}, \bibinfo
  {author} {\bibfnamefont {J.}~\bibnamefont {Manley}}, \bibinfo {author}
  {\bibfnamefont {N.}~\bibnamefont {Matsumoto}}, \bibinfo {author}
  {\bibfnamefont {A.}~\bibnamefont {Monte}}, \bibinfo {author} {\bibfnamefont
  {F.}~\bibnamefont {Monteiro}}, \bibinfo {author} {\bibfnamefont
  {T.}~\bibnamefont {Purdy}}, \bibinfo {author} {\bibfnamefont {C.~J.}\
  \bibnamefont {Riedel}}, \bibinfo {author} {\bibfnamefont {R.}~\bibnamefont
  {Singh}}, \bibinfo {author} {\bibfnamefont {S.}~\bibnamefont {Singh}},
  \bibinfo {author} {\bibfnamefont {K.}~\bibnamefont {Sinha}}, \bibinfo
  {author} {\bibfnamefont {J.~M.}\ \bibnamefont {Taylor}}, \bibinfo {author}
  {\bibfnamefont {J.}~\bibnamefont {Qin}}, \bibinfo {author} {\bibfnamefont
  {D.~J.}\ \bibnamefont {Wilson}}, \ and\ \bibinfo {author} {\bibfnamefont
  {Y.}~\bibnamefont {Zhao}},\ }\href {\doibase 10.1088/2058-9565/abcfcd}
  {\bibfield  {journal} {\bibinfo  {journal} {Quantum Science and Technology}\
  }\textbf {\bibinfo {volume} {6}},\ \bibinfo {pages} {024002} (\bibinfo {year}
  {2021})}\BibitemShut {NoStop}%
\bibitem [{\citenamefont {Monteiro}\ \emph
  {et~al.}(2020{\natexlab{a}})\citenamefont {Monteiro}, \citenamefont {Li},
  \citenamefont {Afek}, \citenamefont {Li}, \citenamefont {Mossman},\ and\
  \citenamefont {Moore}}]{Monteiro2020}%
  \BibitemOpen
  \bibfield  {author} {\bibinfo {author} {\bibfnamefont {F.}~\bibnamefont
  {Monteiro}}, \bibinfo {author} {\bibfnamefont {W.}~\bibnamefont {Li}},
  \bibinfo {author} {\bibfnamefont {G.}~\bibnamefont {Afek}}, \bibinfo {author}
  {\bibfnamefont {C.-l.}\ \bibnamefont {Li}}, \bibinfo {author} {\bibfnamefont
  {M.}~\bibnamefont {Mossman}}, \ and\ \bibinfo {author} {\bibfnamefont
  {D.~C.}\ \bibnamefont {Moore}},\ }\href {\doibase
  10.1103/PhysRevA.101.053835} {\bibfield  {journal} {\bibinfo  {journal}
  {Phys. Rev. A}\ }\textbf {\bibinfo {volume} {101}},\ \bibinfo {pages}
  {053835} (\bibinfo {year} {2020}{\natexlab{a}})}\BibitemShut {NoStop}%
\bibitem [{\citenamefont {Monteiro}\ \emph
  {et~al.}(2020{\natexlab{b}})\citenamefont {Monteiro}, \citenamefont {Afek},
  \citenamefont {Carney}, \citenamefont {Krnjaic}, \citenamefont {Wang},\ and\
  \citenamefont {Moore}}]{Monteiro2020DM}%
  \BibitemOpen
  \bibfield  {author} {\bibinfo {author} {\bibfnamefont {F.}~\bibnamefont
  {Monteiro}}, \bibinfo {author} {\bibfnamefont {G.}~\bibnamefont {Afek}},
  \bibinfo {author} {\bibfnamefont {D.}~\bibnamefont {Carney}}, \bibinfo
  {author} {\bibfnamefont {G.}~\bibnamefont {Krnjaic}}, \bibinfo {author}
  {\bibfnamefont {J.}~\bibnamefont {Wang}}, \ and\ \bibinfo {author}
  {\bibfnamefont {D.~C.}\ \bibnamefont {Moore}},\ }\href {\doibase
  10.1103/PhysRevLett.125.181102} {\bibfield  {journal} {\bibinfo  {journal}
  {Phys. Rev. Lett.}\ }\textbf {\bibinfo {volume} {125}},\ \bibinfo {pages}
  {181102} (\bibinfo {year} {2020}{\natexlab{b}})}\BibitemShut {NoStop}%
\bibitem [{\citenamefont {Safronova}\ \emph {et~al.}(2018)\citenamefont
  {Safronova}, \citenamefont {Budker}, \citenamefont {DeMille}, \citenamefont
  {Kimball}, \citenamefont {Derevianko},\ and\ \citenamefont
  {Clark}}]{Safronova2018}%
  \BibitemOpen
  \bibfield  {author} {\bibinfo {author} {\bibfnamefont {M.~S.}\ \bibnamefont
  {Safronova}}, \bibinfo {author} {\bibfnamefont {D.}~\bibnamefont {Budker}},
  \bibinfo {author} {\bibfnamefont {D.}~\bibnamefont {DeMille}}, \bibinfo
  {author} {\bibfnamefont {D.~F.~J.}\ \bibnamefont {Kimball}}, \bibinfo
  {author} {\bibfnamefont {A.}~\bibnamefont {Derevianko}}, \ and\ \bibinfo
  {author} {\bibfnamefont {C.~W.}\ \bibnamefont {Clark}},\ }\href {\doibase
  10.1103/RevModPhys.90.025008} {\bibfield  {journal} {\bibinfo  {journal}
  {Rev. Mod. Phys.}\ }\textbf {\bibinfo {volume} {90}},\ \bibinfo {pages}
  {025008} (\bibinfo {year} {2018})}\BibitemShut {NoStop}%
\bibitem [{\citenamefont {Parker}\ \emph {et~al.}(2018)\citenamefont {Parker},
  \citenamefont {Yu}, \citenamefont {Zhong}, \citenamefont {Estey},\ and\
  \citenamefont {M{\"u}ller}}]{Parker2018}%
  \BibitemOpen
  \bibfield  {author} {\bibinfo {author} {\bibfnamefont {R.~H.}\ \bibnamefont
  {Parker}}, \bibinfo {author} {\bibfnamefont {C.}~\bibnamefont {Yu}}, \bibinfo
  {author} {\bibfnamefont {W.}~\bibnamefont {Zhong}}, \bibinfo {author}
  {\bibfnamefont {B.}~\bibnamefont {Estey}}, \ and\ \bibinfo {author}
  {\bibfnamefont {H.}~\bibnamefont {M{\"u}ller}},\ }\href {\doibase
  10.1126/science.aap7706} {\bibfield  {journal} {\bibinfo  {journal}
  {Science}\ }\textbf {\bibinfo {volume} {360}},\ \bibinfo {pages} {191}
  (\bibinfo {year} {2018})}\BibitemShut {NoStop}%
\bibitem [{\citenamefont {Morel}\ \emph {et~al.}(2020)\citenamefont {Morel},
  \citenamefont {Yao}, \citenamefont {Clad{\'e}},\ and\ \citenamefont
  {Guellati-Kh{\'e}lifa}}]{Morel2020}%
  \BibitemOpen
  \bibfield  {author} {\bibinfo {author} {\bibfnamefont {L.}~\bibnamefont
  {Morel}}, \bibinfo {author} {\bibfnamefont {Z.}~\bibnamefont {Yao}}, \bibinfo
  {author} {\bibfnamefont {P.}~\bibnamefont {Clad{\'e}}}, \ and\ \bibinfo
  {author} {\bibfnamefont {S.}~\bibnamefont {Guellati-Kh{\'e}lifa}},\
  }\href@noop {} {\bibfield  {journal} {\bibinfo  {journal} {Nature}\ }\textbf
  {\bibinfo {volume} {588}},\ \bibinfo {pages} {61} (\bibinfo {year}
  {2020})}\BibitemShut {NoStop}%
\bibitem [{\citenamefont {Kennedy}\ \emph {et~al.}(2020)\citenamefont
  {Kennedy}, \citenamefont {Oelker}, \citenamefont {Robinson}, \citenamefont
  {Bothwell}, \citenamefont {Kedar}, \citenamefont {Milner}, \citenamefont
  {Marti}, \citenamefont {Derevianko},\ and\ \citenamefont {Ye}}]{Kennedy2020}%
  \BibitemOpen
  \bibfield  {author} {\bibinfo {author} {\bibfnamefont {C.~J.}\ \bibnamefont
  {Kennedy}}, \bibinfo {author} {\bibfnamefont {E.}~\bibnamefont {Oelker}},
  \bibinfo {author} {\bibfnamefont {J.~M.}\ \bibnamefont {Robinson}}, \bibinfo
  {author} {\bibfnamefont {T.}~\bibnamefont {Bothwell}}, \bibinfo {author}
  {\bibfnamefont {D.}~\bibnamefont {Kedar}}, \bibinfo {author} {\bibfnamefont
  {W.~R.}\ \bibnamefont {Milner}}, \bibinfo {author} {\bibfnamefont {G.~E.}\
  \bibnamefont {Marti}}, \bibinfo {author} {\bibfnamefont {A.}~\bibnamefont
  {Derevianko}}, \ and\ \bibinfo {author} {\bibfnamefont {J.}~\bibnamefont
  {Ye}},\ }\href {\doibase 10.1103/PhysRevLett.125.201302} {\bibfield
  {journal} {\bibinfo  {journal} {Phys. Rev. Lett.}\ }\textbf {\bibinfo
  {volume} {125}},\ \bibinfo {pages} {201302} (\bibinfo {year}
  {2020})}\BibitemShut {NoStop}%
\bibitem [{\citenamefont {Moore}\ \emph {et~al.}(2014)\citenamefont {Moore},
  \citenamefont {Rider},\ and\ \citenamefont {Gratta}}]{Moore2014}%
  \BibitemOpen
  \bibfield  {author} {\bibinfo {author} {\bibfnamefont {D.~C.}\ \bibnamefont
  {Moore}}, \bibinfo {author} {\bibfnamefont {A.~D.}\ \bibnamefont {Rider}}, \
  and\ \bibinfo {author} {\bibfnamefont {G.}~\bibnamefont {Gratta}},\ }\href
  {\doibase 10.1103/PhysRevLett.113.251801} {\bibfield  {journal} {\bibinfo
  {journal} {Phys. Rev. Lett.}\ }\textbf {\bibinfo {volume} {113}},\ \bibinfo
  {pages} {251801} (\bibinfo {year} {2014})}\BibitemShut {NoStop}%
\bibitem [{\citenamefont {Monteiro}\ \emph {et~al.}(2017)\citenamefont
  {Monteiro}, \citenamefont {Ghosh}, \citenamefont {Fine},\ and\ \citenamefont
  {Moore}}]{Monteiro2017}%
  \BibitemOpen
  \bibfield  {author} {\bibinfo {author} {\bibfnamefont {F.}~\bibnamefont
  {Monteiro}}, \bibinfo {author} {\bibfnamefont {S.}~\bibnamefont {Ghosh}},
  \bibinfo {author} {\bibfnamefont {A.~G.}\ \bibnamefont {Fine}}, \ and\
  \bibinfo {author} {\bibfnamefont {D.~C.}\ \bibnamefont {Moore}},\ }\href
  {\doibase 10.1103/PhysRevA.96.063841} {\bibfield  {journal} {\bibinfo
  {journal} {Phys. Rev. A}\ }\textbf {\bibinfo {volume} {96}},\ \bibinfo
  {pages} {063841} (\bibinfo {year} {2017})}\BibitemShut {NoStop}%
\bibitem [{\citenamefont {Monteiro}\ \emph {et~al.}(2018)\citenamefont
  {Monteiro}, \citenamefont {Ghosh}, \citenamefont {van Assendelft},\ and\
  \citenamefont {Moore}}]{Monteiro2018}%
  \BibitemOpen
  \bibfield  {author} {\bibinfo {author} {\bibfnamefont {F.}~\bibnamefont
  {Monteiro}}, \bibinfo {author} {\bibfnamefont {S.}~\bibnamefont {Ghosh}},
  \bibinfo {author} {\bibfnamefont {E.~C.}\ \bibnamefont {van Assendelft}}, \
  and\ \bibinfo {author} {\bibfnamefont {D.~C.}\ \bibnamefont {Moore}},\ }\href
  {\doibase 10.1103/PhysRevA.97.051802} {\bibfield  {journal} {\bibinfo
  {journal} {Phys. Rev. A}\ }\textbf {\bibinfo {volume} {97}},\ \bibinfo
  {pages} {051802} (\bibinfo {year} {2018})}\BibitemShut {NoStop}%
\bibitem [{\citenamefont {Bressi}\ \emph {et~al.}(2011)\citenamefont {Bressi},
  \citenamefont {Carugno}, \citenamefont {Della~Valle}, \citenamefont
  {Galeazzi}, \citenamefont {Ruoso},\ and\ \citenamefont
  {Sartori}}]{Bressi2011}%
  \BibitemOpen
  \bibfield  {author} {\bibinfo {author} {\bibfnamefont {G.}~\bibnamefont
  {Bressi}}, \bibinfo {author} {\bibfnamefont {G.}~\bibnamefont {Carugno}},
  \bibinfo {author} {\bibfnamefont {F.}~\bibnamefont {Della~Valle}}, \bibinfo
  {author} {\bibfnamefont {G.}~\bibnamefont {Galeazzi}}, \bibinfo {author}
  {\bibfnamefont {G.}~\bibnamefont {Ruoso}}, \ and\ \bibinfo {author}
  {\bibfnamefont {G.}~\bibnamefont {Sartori}},\ }\href {\doibase
  10.1103/PhysRevA.83.052101} {\bibfield  {journal} {\bibinfo  {journal} {Phys.
  Rev. A}\ }\textbf {\bibinfo {volume} {83}},\ \bibinfo {pages} {052101}
  (\bibinfo {year} {2011})}\BibitemShut {NoStop}%
\bibitem [{\citenamefont {Baumann}\ \emph {et~al.}(1988)\citenamefont
  {Baumann}, \citenamefont {G\"ahler}, \citenamefont {Kalus},\ and\
  \citenamefont {Mampe}}]{Baumann1988}%
  \BibitemOpen
  \bibfield  {author} {\bibinfo {author} {\bibfnamefont {J.}~\bibnamefont
  {Baumann}}, \bibinfo {author} {\bibfnamefont {R.}~\bibnamefont {G\"ahler}},
  \bibinfo {author} {\bibfnamefont {J.}~\bibnamefont {Kalus}}, \ and\ \bibinfo
  {author} {\bibfnamefont {W.}~\bibnamefont {Mampe}},\ }\href {\doibase
  10.1103/PhysRevD.37.3107} {\bibfield  {journal} {\bibinfo  {journal} {Phys.
  Rev. D}\ }\textbf {\bibinfo {volume} {37}},\ \bibinfo {pages} {3107}
  (\bibinfo {year} {1988})}\BibitemShut {NoStop}%
\bibitem [{\citenamefont {Davidson}\ \emph {et~al.}(2000)\citenamefont
  {Davidson}, \citenamefont {Hannestad},\ and\ \citenamefont
  {Raffelt}}]{Davidson2000LEP}%
  \BibitemOpen
  \bibfield  {author} {\bibinfo {author} {\bibfnamefont {S.}~\bibnamefont
  {Davidson}}, \bibinfo {author} {\bibfnamefont {S.}~\bibnamefont {Hannestad}},
  \ and\ \bibinfo {author} {\bibfnamefont {G.}~\bibnamefont {Raffelt}},\ }\href
  {\doibase 10.1088/1126-6708/2000/05/003} {\bibfield  {journal} {\bibinfo
  {journal} {JHEP}\ }\textbf {\bibinfo {volume} {2000}},\ \bibinfo {pages}
  {003} (\bibinfo {year} {2000})}\BibitemShut {NoStop}%
\bibitem [{\citenamefont {Jaeckel}\ \emph {et~al.}(2013)\citenamefont
  {Jaeckel}, \citenamefont {Jankowiak},\ and\ \citenamefont
  {Spannowsky}}]{Jaeckel2013LHC}%
  \BibitemOpen
  \bibfield  {author} {\bibinfo {author} {\bibfnamefont {J.}~\bibnamefont
  {Jaeckel}}, \bibinfo {author} {\bibfnamefont {M.}~\bibnamefont {Jankowiak}},
  \ and\ \bibinfo {author} {\bibfnamefont {M.}~\bibnamefont {Spannowsky}},\
  }\href {\doibase https://doi.org/10.1016/j.dark.2013.06.001} {\bibfield
  {journal} {\bibinfo  {journal} {Phys. Dark Universe}\ }\textbf {\bibinfo
  {volume} {2}},\ \bibinfo {pages} {111 } (\bibinfo {year} {2013})}\BibitemShut
  {NoStop}%
\bibitem [{\citenamefont {Chatrchyan}\ \emph
  {et~al.}(2013{\natexlab{a}})\citenamefont {Chatrchyan} \emph
  {et~al.}}]{CMS:2012xi}%
  \BibitemOpen
  \bibfield  {author} {\bibinfo {author} {\bibfnamefont {S.}~\bibnamefont
  {Chatrchyan}} \emph {et~al.} (\bibinfo {collaboration} {CMS}),\ }\href
  {\doibase 10.1103/PhysRevD.87.092008} {\bibfield  {journal} {\bibinfo
  {journal} {Phys. Rev. D}\ }\textbf {\bibinfo {volume} {87}},\ \bibinfo
  {pages} {092008} (\bibinfo {year} {2013}{\natexlab{a}})},\ \Eprint
  {http://arxiv.org/abs/1210.2311} {arXiv:1210.2311 [hep-ex]} \BibitemShut
  {NoStop}%
\bibitem [{\citenamefont {Chatrchyan}\ \emph
  {et~al.}(2013{\natexlab{b}})\citenamefont {Chatrchyan} \emph
  {et~al.}}]{Chatrchyan:2013oca}%
  \BibitemOpen
  \bibfield  {author} {\bibinfo {author} {\bibfnamefont {S.}~\bibnamefont
  {Chatrchyan}} \emph {et~al.} (\bibinfo {collaboration} {CMS}),\ }\href
  {\doibase 10.1007/JHEP07(2013)122} {\bibfield  {journal} {\bibinfo  {journal}
  {JHEP}\ }\textbf {\bibinfo {volume} {07}},\ \bibinfo {pages} {122} (\bibinfo
  {year} {2013}{\natexlab{b}})},\ \Eprint {http://arxiv.org/abs/1305.0491}
  {arXiv:1305.0491 [hep-ex]} \BibitemShut {NoStop}%
\bibitem [{\citenamefont {Acciarri}\ \emph {et~al.}(2020)\citenamefont
  {Acciarri} \emph {et~al.}}]{Acciarri2020ArgoNeuT}%
  \BibitemOpen
  \bibfield  {author} {\bibinfo {author} {\bibfnamefont {R.}~\bibnamefont
  {Acciarri}} \emph {et~al.} (\bibinfo {collaboration} {ArgoNeuT
  Collaboration}),\ }\href {\doibase 10.1103/PhysRevLett.124.131801} {\bibfield
   {journal} {\bibinfo  {journal} {Phys. Rev. Lett.}\ }\textbf {\bibinfo
  {volume} {124}},\ \bibinfo {pages} {131801} (\bibinfo {year}
  {2020})}\BibitemShut {NoStop}%
\bibitem [{\citenamefont {Magill}\ \emph {et~al.}(2019)\citenamefont {Magill},
  \citenamefont {Plestid}, \citenamefont {Pospelov},\ and\ \citenamefont
  {Tsai}}]{Magill:2018tbb}%
  \BibitemOpen
  \bibfield  {author} {\bibinfo {author} {\bibfnamefont {G.}~\bibnamefont
  {Magill}}, \bibinfo {author} {\bibfnamefont {R.}~\bibnamefont {Plestid}},
  \bibinfo {author} {\bibfnamefont {M.}~\bibnamefont {Pospelov}}, \ and\
  \bibinfo {author} {\bibfnamefont {Y.-D.}\ \bibnamefont {Tsai}},\ }\href
  {\doibase 10.1103/PhysRevLett.122.071801} {\bibfield  {journal} {\bibinfo
  {journal} {Phys. Rev. Lett.}\ }\textbf {\bibinfo {volume} {122}},\ \bibinfo
  {pages} {071801} (\bibinfo {year} {2019})},\ \Eprint
  {http://arxiv.org/abs/1806.03310} {arXiv:1806.03310 [hep-ph]} \BibitemShut
  {NoStop}%
\bibitem [{\citenamefont {Prinz}\ \emph {et~al.}(1998)\citenamefont {Prinz}
  \emph {et~al.}}]{Prinz1998SLAC}%
  \BibitemOpen
  \bibfield  {author} {\bibinfo {author} {\bibfnamefont {A.~A.}\ \bibnamefont
  {Prinz}} \emph {et~al.},\ }\href {\doibase 10.1103/PhysRevLett.81.1175}
  {\bibfield  {journal} {\bibinfo  {journal} {Phys. Rev. Lett.}\ }\textbf
  {\bibinfo {volume} {81}},\ \bibinfo {pages} {1175} (\bibinfo {year}
  {1998})}\BibitemShut {NoStop}%
\bibitem [{\citenamefont {Ball}\ \emph {et~al.}(2020)\citenamefont {Ball} \emph
  {et~al.}}]{Ball2020MilliquanDemo}%
  \BibitemOpen
  \bibfield  {author} {\bibinfo {author} {\bibfnamefont {A.}~\bibnamefont
  {Ball}} \emph {et~al.},\ }\href {\doibase 10.1103/PhysRevD.102.032002}
  {\bibfield  {journal} {\bibinfo  {journal} {Phys. Rev. D}\ }\textbf {\bibinfo
  {volume} {102}},\ \bibinfo {pages} {032002} (\bibinfo {year}
  {2020})}\BibitemShut {NoStop}%
\bibitem [{\citenamefont {Plestid}\ \emph {et~al.}(2020)\citenamefont
  {Plestid}, \citenamefont {Takhistov}, \citenamefont {Tsai}, \citenamefont
  {Bringmann}, \citenamefont {Kusenko},\ and\ \citenamefont
  {Pospelov}}]{Plestid:2020kdm}%
  \BibitemOpen
  \bibfield  {author} {\bibinfo {author} {\bibfnamefont {R.}~\bibnamefont
  {Plestid}}, \bibinfo {author} {\bibfnamefont {V.}~\bibnamefont {Takhistov}},
  \bibinfo {author} {\bibfnamefont {Y.-D.}\ \bibnamefont {Tsai}}, \bibinfo
  {author} {\bibfnamefont {T.}~\bibnamefont {Bringmann}}, \bibinfo {author}
  {\bibfnamefont {A.}~\bibnamefont {Kusenko}}, \ and\ \bibinfo {author}
  {\bibfnamefont {M.}~\bibnamefont {Pospelov}},\ }\href@noop {} {\bibfield
  {journal} {\bibinfo  {journal} {arXiv preprint arXiv:2002.11732}\ } (\bibinfo
  {year} {2020})}\BibitemShut {NoStop}%
\bibitem [{\citenamefont {Alvis}\ \emph {et~al.}(2018)\citenamefont {Alvis}
  \emph {et~al.}}]{Alvis:2018yte}%
  \BibitemOpen
  \bibfield  {author} {\bibinfo {author} {\bibfnamefont {S.}~\bibnamefont
  {Alvis}} \emph {et~al.} (\bibinfo {collaboration} {Majorana}),\ }\href
  {\doibase 10.1103/PhysRevLett.120.211804} {\bibfield  {journal} {\bibinfo
  {journal} {Phys. Rev. Lett.}\ }\textbf {\bibinfo {volume} {120}},\ \bibinfo
  {pages} {211804} (\bibinfo {year} {2018})},\ \Eprint
  {http://arxiv.org/abs/1801.10145} {arXiv:1801.10145 [hep-ex]} \BibitemShut
  {NoStop}%
\bibitem [{\citenamefont {Singh}\ \emph {et~al.}(2019)\citenamefont {Singh}
  \emph {et~al.}}]{Singh:2018von}%
  \BibitemOpen
  \bibfield  {author} {\bibinfo {author} {\bibfnamefont {L.}~\bibnamefont
  {Singh}} \emph {et~al.} (\bibinfo {collaboration} {TEXONO}),\ }\href
  {\doibase 10.1103/PhysRevD.99.032009} {\bibfield  {journal} {\bibinfo
  {journal} {Phys. Rev. D}\ }\textbf {\bibinfo {volume} {99}},\ \bibinfo
  {pages} {032009} (\bibinfo {year} {2019})},\ \Eprint
  {http://arxiv.org/abs/1808.02719} {arXiv:1808.02719 [hep-ph]} \BibitemShut
  {NoStop}%
\bibitem [{\citenamefont {Alkhatib}\ \emph {et~al.}(2020)\citenamefont
  {Alkhatib}, \citenamefont {Amaral}, \citenamefont {Aralis}, \citenamefont
  {Aramaki}, \citenamefont {Arnquist}, \citenamefont {Langroudy}, \citenamefont
  {Azadbakht}, \citenamefont {Banik}, \citenamefont {Barker}, \citenamefont
  {Bathurst} \emph {et~al.}}]{alkhatib2020constraints}%
  \BibitemOpen
  \bibfield  {author} {\bibinfo {author} {\bibfnamefont {I.}~\bibnamefont
  {Alkhatib}}, \bibinfo {author} {\bibfnamefont {D.}~\bibnamefont {Amaral}},
  \bibinfo {author} {\bibfnamefont {T.}~\bibnamefont {Aralis}}, \bibinfo
  {author} {\bibfnamefont {T.}~\bibnamefont {Aramaki}}, \bibinfo {author}
  {\bibfnamefont {I.}~\bibnamefont {Arnquist}}, \bibinfo {author}
  {\bibfnamefont {I.~A.}\ \bibnamefont {Langroudy}}, \bibinfo {author}
  {\bibfnamefont {E.}~\bibnamefont {Azadbakht}}, \bibinfo {author}
  {\bibfnamefont {S.}~\bibnamefont {Banik}}, \bibinfo {author} {\bibfnamefont
  {D.}~\bibnamefont {Barker}}, \bibinfo {author} {\bibfnamefont
  {C.}~\bibnamefont {Bathurst}},  \emph {et~al.},\ }\href@noop {} {\bibfield
  {journal} {\bibinfo  {journal} {arXiv preprint arXiv:2011.09183}\ } (\bibinfo
  {year} {2020})}\BibitemShut {NoStop}%
\bibitem [{\citenamefont {Agnese}\ \emph {et~al.}(2015)\citenamefont {Agnese}
  \emph {et~al.}}]{Agnese:2014vxh}%
  \BibitemOpen
  \bibfield  {author} {\bibinfo {author} {\bibfnamefont {R.}~\bibnamefont
  {Agnese}} \emph {et~al.} (\bibinfo {collaboration} {CDMS}),\ }\href {\doibase
  10.1103/PhysRevLett.114.111302} {\bibfield  {journal} {\bibinfo  {journal}
  {Phys. Rev. Lett.}\ }\textbf {\bibinfo {volume} {114}},\ \bibinfo {pages}
  {111302} (\bibinfo {year} {2015})},\ \Eprint {http://arxiv.org/abs/1409.3270}
  {arXiv:1409.3270 [hep-ex]} \BibitemShut {NoStop}%
\bibitem [{\citenamefont {Pospelov}\ and\ \citenamefont
  {Ramani}(2020)}]{pospelov2020earthbound}%
  \BibitemOpen
  \bibfield  {author} {\bibinfo {author} {\bibfnamefont {M.}~\bibnamefont
  {Pospelov}}\ and\ \bibinfo {author} {\bibfnamefont {H.}~\bibnamefont
  {Ramani}},\ }\href@noop {} {\bibfield  {journal} {\bibinfo  {journal} {arXiv
  preprint arXiv:2012.03957}\ } (\bibinfo {year} {2020})},\ \Eprint
  {http://arxiv.org/abs/2012.03957} {arXiv:2012.03957 [hep-ph]} \BibitemShut
  {NoStop}%
\bibitem [{\citenamefont {Langacker}\ and\ \citenamefont
  {Steigman}(2011)}]{Langacker2011}%
  \BibitemOpen
  \bibfield  {author} {\bibinfo {author} {\bibfnamefont {P.}~\bibnamefont
  {Langacker}}\ and\ \bibinfo {author} {\bibfnamefont {G.}~\bibnamefont
  {Steigman}},\ }\href {\doibase 10.1103/PhysRevD.84.065040} {\bibfield
  {journal} {\bibinfo  {journal} {Phys. Rev. D}\ }\textbf {\bibinfo {volume}
  {84}},\ \bibinfo {pages} {065040} (\bibinfo {year} {2011})}\BibitemShut
  {NoStop}%
\bibitem [{\citenamefont {LaRue}\ \emph {et~al.}(1981)\citenamefont {LaRue},
  \citenamefont {Phillips},\ and\ \citenamefont {Fairbank}}]{LaRue1981}%
  \BibitemOpen
  \bibfield  {author} {\bibinfo {author} {\bibfnamefont {G.~S.}\ \bibnamefont
  {LaRue}}, \bibinfo {author} {\bibfnamefont {J.~D.}\ \bibnamefont {Phillips}},
  \ and\ \bibinfo {author} {\bibfnamefont {W.~M.}\ \bibnamefont {Fairbank}},\
  }\href {\doibase 10.1103/PhysRevLett.46.967} {\bibfield  {journal} {\bibinfo
  {journal} {Phys. Rev. Lett.}\ }\textbf {\bibinfo {volume} {46}},\ \bibinfo
  {pages} {967} (\bibinfo {year} {1981})}\BibitemShut {NoStop}%
\bibitem [{\citenamefont {Marinelli}\ and\ \citenamefont
  {Morpurgo}(1982)}]{Marinelli1982}%
  \BibitemOpen
  \bibfield  {author} {\bibinfo {author} {\bibfnamefont {M.}~\bibnamefont
  {Marinelli}}\ and\ \bibinfo {author} {\bibfnamefont {G.}~\bibnamefont
  {Morpurgo}},\ }\href {\doibase https://doi.org/10.1016/0370-1573(82)90053-9}
  {\bibfield  {journal} {\bibinfo  {journal} {Physics Reports}\ }\textbf
  {\bibinfo {volume} {85}},\ \bibinfo {pages} {161 } (\bibinfo {year}
  {1982})}\BibitemShut {NoStop}%
\bibitem [{\citenamefont {Phillips}\ \emph {et~al.}(1988)\citenamefont
  {Phillips}, \citenamefont {Fairbank},\ and\ \citenamefont
  {Navarro}}]{Phillips1988}%
  \BibitemOpen
  \bibfield  {author} {\bibinfo {author} {\bibfnamefont {J.~D.}\ \bibnamefont
  {Phillips}}, \bibinfo {author} {\bibfnamefont {W.~M.}\ \bibnamefont
  {Fairbank}}, \ and\ \bibinfo {author} {\bibfnamefont {J.}~\bibnamefont
  {Navarro}},\ }\href {\doibase https://doi.org/10.1016/0168-9002(88)91113-8}
  {\bibfield  {journal} {\bibinfo  {journal} {Nucl. Instrum. Meth. Phys. Res.
  A}\ }\textbf {\bibinfo {volume} {264}},\ \bibinfo {pages} {125 } (\bibinfo
  {year} {1988})}\BibitemShut {NoStop}%
\bibitem [{\citenamefont {Smith}(1989)}]{Smith1989}%
  \BibitemOpen
  \bibfield  {author} {\bibinfo {author} {\bibfnamefont {P.~F.}\ \bibnamefont
  {Smith}},\ }\href {\doibase 10.1146/annurev.ns.39.120189.000445} {\bibfield
  {journal} {\bibinfo  {journal} {Ann. Rev. Nucl. Part. Sci.}\ }\textbf
  {\bibinfo {volume} {39}},\ \bibinfo {pages} {73} (\bibinfo {year}
  {1989})}\BibitemShut {NoStop}%
\bibitem [{\citenamefont {Joyce}\ \emph {et~al.}(1983)\citenamefont {Joyce},
  \citenamefont {Abrams}, \citenamefont {Bland}, \citenamefont {Johnson},
  \citenamefont {Lindgren}, \citenamefont {Savage}, \citenamefont {Scholz},
  \citenamefont {Young},\ and\ \citenamefont {Hodges}}]{Joyce1983}%
  \BibitemOpen
  \bibfield  {author} {\bibinfo {author} {\bibfnamefont {D.~C.}\ \bibnamefont
  {Joyce}}, \bibinfo {author} {\bibfnamefont {P.~C.}\ \bibnamefont {Abrams}},
  \bibinfo {author} {\bibfnamefont {R.~W.}\ \bibnamefont {Bland}}, \bibinfo
  {author} {\bibfnamefont {R.~T.}\ \bibnamefont {Johnson}}, \bibinfo {author}
  {\bibfnamefont {M.~A.}\ \bibnamefont {Lindgren}}, \bibinfo {author}
  {\bibfnamefont {M.~H.}\ \bibnamefont {Savage}}, \bibinfo {author}
  {\bibfnamefont {M.~H.}\ \bibnamefont {Scholz}}, \bibinfo {author}
  {\bibfnamefont {B.~A.}\ \bibnamefont {Young}}, \ and\ \bibinfo {author}
  {\bibfnamefont {C.~L.}\ \bibnamefont {Hodges}},\ }\href {\doibase
  10.1103/PhysRevLett.51.731} {\bibfield  {journal} {\bibinfo  {journal} {Phys.
  Rev. Lett.}\ }\textbf {\bibinfo {volume} {51}},\ \bibinfo {pages} {731}
  (\bibinfo {year} {1983})}\BibitemShut {NoStop}%
\bibitem [{\citenamefont {Van~Polen}\ \emph {et~al.}(1987)\citenamefont
  {Van~Polen}, \citenamefont {Hagstrom},\ and\ \citenamefont
  {Hirsch}}]{VanPolen1987}%
  \BibitemOpen
  \bibfield  {author} {\bibinfo {author} {\bibfnamefont {J.}~\bibnamefont
  {Van~Polen}}, \bibinfo {author} {\bibfnamefont {R.~T.}\ \bibnamefont
  {Hagstrom}}, \ and\ \bibinfo {author} {\bibfnamefont {G.}~\bibnamefont
  {Hirsch}},\ }\href {\doibase 10.1103/PhysRevD.36.1983} {\bibfield  {journal}
  {\bibinfo  {journal} {Phys. Rev. D}\ }\textbf {\bibinfo {volume} {36}},\
  \bibinfo {pages} {1983} (\bibinfo {year} {1987})}\BibitemShut {NoStop}%
\bibitem [{\citenamefont {Lee}\ \emph {et~al.}(2002)\citenamefont {Lee},
  \citenamefont {Fan}, \citenamefont {Halyo}, \citenamefont {Lee},
  \citenamefont {Kim}, \citenamefont {Perl}, \citenamefont {Rogers},
  \citenamefont {Loomba}, \citenamefont {Lackner},\ and\ \citenamefont
  {Shaw}}]{Lee2002}%
  \BibitemOpen
  \bibfield  {author} {\bibinfo {author} {\bibfnamefont {I.~T.}\ \bibnamefont
  {Lee}}, \bibinfo {author} {\bibfnamefont {S.}~\bibnamefont {Fan}}, \bibinfo
  {author} {\bibfnamefont {V.}~\bibnamefont {Halyo}}, \bibinfo {author}
  {\bibfnamefont {E.~R.}\ \bibnamefont {Lee}}, \bibinfo {author} {\bibfnamefont
  {P.~C.}\ \bibnamefont {Kim}}, \bibinfo {author} {\bibfnamefont {M.~L.}\
  \bibnamefont {Perl}}, \bibinfo {author} {\bibfnamefont {H.}~\bibnamefont
  {Rogers}}, \bibinfo {author} {\bibfnamefont {D.}~\bibnamefont {Loomba}},
  \bibinfo {author} {\bibfnamefont {K.~S.}\ \bibnamefont {Lackner}}, \ and\
  \bibinfo {author} {\bibfnamefont {G.}~\bibnamefont {Shaw}},\ }\href {\doibase
  10.1103/PhysRevD.66.012002} {\bibfield  {journal} {\bibinfo  {journal} {Phys.
  Rev. D}\ }\textbf {\bibinfo {volume} {66}},\ \bibinfo {pages} {012002}
  (\bibinfo {year} {2002})}\BibitemShut {NoStop}%
\bibitem [{\citenamefont {Kim}\ \emph {et~al.}(2007)\citenamefont {Kim},
  \citenamefont {Lee}, \citenamefont {Lee}, \citenamefont {Perl}, \citenamefont
  {Halyo},\ and\ \citenamefont {Loomba}}]{Kim2007}%
  \BibitemOpen
  \bibfield  {author} {\bibinfo {author} {\bibfnamefont {P.~C.}\ \bibnamefont
  {Kim}}, \bibinfo {author} {\bibfnamefont {E.~R.}\ \bibnamefont {Lee}},
  \bibinfo {author} {\bibfnamefont {I.~T.}\ \bibnamefont {Lee}}, \bibinfo
  {author} {\bibfnamefont {M.~L.}\ \bibnamefont {Perl}}, \bibinfo {author}
  {\bibfnamefont {V.}~\bibnamefont {Halyo}}, \ and\ \bibinfo {author}
  {\bibfnamefont {D.}~\bibnamefont {Loomba}},\ }\href {\doibase
  10.1103/PhysRevLett.99.161804} {\bibfield  {journal} {\bibinfo  {journal}
  {Phys. Rev. Lett.}\ }\textbf {\bibinfo {volume} {99}},\ \bibinfo {pages}
  {161804} (\bibinfo {year} {2007})}\BibitemShut {NoStop}%
\bibitem [{Note1()}]{Note1}%
  \BibitemOpen
  \bibinfo {note} {\protect \url
  {https://www.microspheres-nanospheres.com/}}\BibitemShut {NoStop}%
\bibitem [{\citenamefont {Frimmer}\ \emph {et~al.}(2017)\citenamefont
  {Frimmer}, \citenamefont {Luszcz}, \citenamefont {Ferreiro}, \citenamefont
  {Jain}, \citenamefont {Hebestreit},\ and\ \citenamefont
  {Novotny}}]{Frimmer2017}%
  \BibitemOpen
  \bibfield  {author} {\bibinfo {author} {\bibfnamefont {M.}~\bibnamefont
  {Frimmer}}, \bibinfo {author} {\bibfnamefont {K.}~\bibnamefont {Luszcz}},
  \bibinfo {author} {\bibfnamefont {S.}~\bibnamefont {Ferreiro}}, \bibinfo
  {author} {\bibfnamefont {V.}~\bibnamefont {Jain}}, \bibinfo {author}
  {\bibfnamefont {E.}~\bibnamefont {Hebestreit}}, \ and\ \bibinfo {author}
  {\bibfnamefont {L.}~\bibnamefont {Novotny}},\ }\href {\doibase
  10.1103/PhysRevA.95.061801} {\bibfield  {journal} {\bibinfo  {journal} {Phys.
  Rev. A}\ }\textbf {\bibinfo {volume} {95}},\ \bibinfo {pages} {061801}
  (\bibinfo {year} {2017})}\BibitemShut {NoStop}%
\bibitem [{\citenamefont {Conangla}\ \emph {et~al.}(2019)\citenamefont
  {Conangla}, \citenamefont {Ricci}, \citenamefont {Cuairan}, \citenamefont
  {Schell}, \citenamefont {Meyer},\ and\ \citenamefont
  {Quidant}}]{Conangla:2018nnn}%
  \BibitemOpen
  \bibfield  {author} {\bibinfo {author} {\bibfnamefont {G.~P.}\ \bibnamefont
  {Conangla}}, \bibinfo {author} {\bibfnamefont {F.}~\bibnamefont {Ricci}},
  \bibinfo {author} {\bibfnamefont {M.~T.}\ \bibnamefont {Cuairan}}, \bibinfo
  {author} {\bibfnamefont {A.~W.}\ \bibnamefont {Schell}}, \bibinfo {author}
  {\bibfnamefont {N.}~\bibnamefont {Meyer}}, \ and\ \bibinfo {author}
  {\bibfnamefont {R.}~\bibnamefont {Quidant}},\ }\href {\doibase
  10.1103/PhysRevLett.122.223602} {\bibfield  {journal} {\bibinfo  {journal}
  {Phys. Rev. Lett.}\ }\textbf {\bibinfo {volume} {122}},\ \bibinfo {pages}
  {223602} (\bibinfo {year} {2019})},\ \Eprint
  {http://arxiv.org/abs/1901.00923} {arXiv:1901.00923 [physics.ins-det]}
  \BibitemShut {NoStop}%
\bibitem [{\citenamefont {{Arita}}\ \emph {et~al.}(2013)\citenamefont
  {{Arita}}, \citenamefont {{Mazilu}},\ and\ \citenamefont
  {{Dholakia}}}]{Arita2013NatCo...4.2374A}%
  \BibitemOpen
  \bibfield  {author} {\bibinfo {author} {\bibfnamefont {Y.}~\bibnamefont
  {{Arita}}}, \bibinfo {author} {\bibfnamefont {M.}~\bibnamefont {{Mazilu}}}, \
  and\ \bibinfo {author} {\bibfnamefont {K.}~\bibnamefont {{Dholakia}}},\
  }\href {\doibase 10.1038/ncomms3374} {\bibfield  {journal} {\bibinfo
  {journal} {Nature Communications}\ }\textbf {\bibinfo {volume} {4}},\
  \bibinfo {eid} {2374} (\bibinfo {year} {2013})}\BibitemShut {NoStop}%
\bibitem [{\citenamefont {Ahn}\ \emph {et~al.}(2018)\citenamefont {Ahn},
  \citenamefont {Xu}, \citenamefont {Bang}, \citenamefont {Deng}, \citenamefont
  {Hoang}, \citenamefont {Han}, \citenamefont {Ma},\ and\ \citenamefont
  {Li}}]{Ahn2018}%
  \BibitemOpen
  \bibfield  {author} {\bibinfo {author} {\bibfnamefont {J.}~\bibnamefont
  {Ahn}}, \bibinfo {author} {\bibfnamefont {Z.}~\bibnamefont {Xu}}, \bibinfo
  {author} {\bibfnamefont {J.}~\bibnamefont {Bang}}, \bibinfo {author}
  {\bibfnamefont {Y.-H.}\ \bibnamefont {Deng}}, \bibinfo {author}
  {\bibfnamefont {T.~M.}\ \bibnamefont {Hoang}}, \bibinfo {author}
  {\bibfnamefont {Q.}~\bibnamefont {Han}}, \bibinfo {author} {\bibfnamefont
  {R.-M.}\ \bibnamefont {Ma}}, \ and\ \bibinfo {author} {\bibfnamefont
  {T.}~\bibnamefont {Li}},\ }\href {\doibase 10.1103/PhysRevLett.121.033603}
  {\bibfield  {journal} {\bibinfo  {journal} {Phys. Rev. Lett.}\ }\textbf
  {\bibinfo {volume} {121}},\ \bibinfo {pages} {033603} (\bibinfo {year}
  {2018})}\BibitemShut {NoStop}%
\bibitem [{\citenamefont {Reimann}\ \emph {et~al.}(2018)\citenamefont
  {Reimann}, \citenamefont {Doderer}, \citenamefont {Hebestreit}, \citenamefont
  {Diehl}, \citenamefont {Frimmer}, \citenamefont {Windey}, \citenamefont
  {Tebbenjohanns},\ and\ \citenamefont {Novotny}}]{Reimann2018}%
  \BibitemOpen
  \bibfield  {author} {\bibinfo {author} {\bibfnamefont {R.}~\bibnamefont
  {Reimann}}, \bibinfo {author} {\bibfnamefont {M.}~\bibnamefont {Doderer}},
  \bibinfo {author} {\bibfnamefont {E.}~\bibnamefont {Hebestreit}}, \bibinfo
  {author} {\bibfnamefont {R.}~\bibnamefont {Diehl}}, \bibinfo {author}
  {\bibfnamefont {M.}~\bibnamefont {Frimmer}}, \bibinfo {author} {\bibfnamefont
  {D.}~\bibnamefont {Windey}}, \bibinfo {author} {\bibfnamefont
  {F.}~\bibnamefont {Tebbenjohanns}}, \ and\ \bibinfo {author} {\bibfnamefont
  {L.}~\bibnamefont {Novotny}},\ }\href {\doibase
  10.1103/PhysRevLett.121.033602} {\bibfield  {journal} {\bibinfo  {journal}
  {Phys. Rev. Lett.}\ }\textbf {\bibinfo {volume} {121}},\ \bibinfo {pages}
  {033602} (\bibinfo {year} {2018})}\BibitemShut {NoStop}%
\bibitem [{\citenamefont {Jackson}(1999)}]{Jackson1999}%
  \BibitemOpen
  \bibfield  {author} {\bibinfo {author} {\bibfnamefont {J.~D.}\ \bibnamefont
  {Jackson}},\ }\href@noop {} {\enquote {\bibinfo {title} {Classical
  electrodynamics},}\ } (\bibinfo {year} {1999})\BibitemShut {NoStop}%
\bibitem [{\citenamefont {Rider}\ \emph {et~al.}(2016)\citenamefont {Rider},
  \citenamefont {Moore}, \citenamefont {Blakemore}, \citenamefont {Louis},
  \citenamefont {Lu},\ and\ \citenamefont {Gratta}}]{Rider:2016xaq}%
  \BibitemOpen
  \bibfield  {author} {\bibinfo {author} {\bibfnamefont {A.~D.}\ \bibnamefont
  {Rider}}, \bibinfo {author} {\bibfnamefont {D.~C.}\ \bibnamefont {Moore}},
  \bibinfo {author} {\bibfnamefont {C.~P.}\ \bibnamefont {Blakemore}}, \bibinfo
  {author} {\bibfnamefont {M.}~\bibnamefont {Louis}}, \bibinfo {author}
  {\bibfnamefont {M.}~\bibnamefont {Lu}}, \ and\ \bibinfo {author}
  {\bibfnamefont {G.}~\bibnamefont {Gratta}},\ }\href {\doibase
  10.1103/PhysRevLett.117.101101} {\bibfield  {journal} {\bibinfo  {journal}
  {Phys. Rev. Lett.}\ }\textbf {\bibinfo {volume} {117}},\ \bibinfo {pages}
  {101101} (\bibinfo {year} {2016})},\ \Eprint
  {http://arxiv.org/abs/1604.04908} {arXiv:1604.04908 [hep-ex]} \BibitemShut
  {NoStop}%
\bibitem [{\citenamefont {Rider}\ \emph {et~al.}(2019)\citenamefont {Rider},
  \citenamefont {Blakemore}, \citenamefont {Kawasaki}, \citenamefont {Priel},
  \citenamefont {Roy},\ and\ \citenamefont
  {Gratta}}]{PhysRevA.99.041802_electric_rotation}%
  \BibitemOpen
  \bibfield  {author} {\bibinfo {author} {\bibfnamefont {A.~D.}\ \bibnamefont
  {Rider}}, \bibinfo {author} {\bibfnamefont {C.~P.}\ \bibnamefont
  {Blakemore}}, \bibinfo {author} {\bibfnamefont {A.}~\bibnamefont {Kawasaki}},
  \bibinfo {author} {\bibfnamefont {N.}~\bibnamefont {Priel}}, \bibinfo
  {author} {\bibfnamefont {S.}~\bibnamefont {Roy}}, \ and\ \bibinfo {author}
  {\bibfnamefont {G.}~\bibnamefont {Gratta}},\ }\href {\doibase
  10.1103/PhysRevA.99.041802} {\bibfield  {journal} {\bibinfo  {journal} {Phys.
  Rev. A}\ }\textbf {\bibinfo {volume} {99}},\ \bibinfo {pages} {041802}
  (\bibinfo {year} {2019})}\BibitemShut {NoStop}%
\bibitem [{\citenamefont {Carretero-Genevrier}\ \emph
  {et~al.}(2013)\citenamefont {Carretero-Genevrier}, \citenamefont {Gich},
  \citenamefont {Picas}, \citenamefont {Gazquez}, \citenamefont {Drisko},
  \citenamefont {Boissiere}, \citenamefont {Grosso}, \citenamefont
  {Rodriguez-Carvajal},\ and\ \citenamefont {Sanchez}}]{Genevrier2013}%
  \BibitemOpen
  \bibfield  {author} {\bibinfo {author} {\bibfnamefont {A.}~\bibnamefont
  {Carretero-Genevrier}}, \bibinfo {author} {\bibfnamefont {M.}~\bibnamefont
  {Gich}}, \bibinfo {author} {\bibfnamefont {L.}~\bibnamefont {Picas}},
  \bibinfo {author} {\bibfnamefont {J.}~\bibnamefont {Gazquez}}, \bibinfo
  {author} {\bibfnamefont {G.~L.}\ \bibnamefont {Drisko}}, \bibinfo {author}
  {\bibfnamefont {C.}~\bibnamefont {Boissiere}}, \bibinfo {author}
  {\bibfnamefont {D.}~\bibnamefont {Grosso}}, \bibinfo {author} {\bibfnamefont
  {J.}~\bibnamefont {Rodriguez-Carvajal}}, \ and\ \bibinfo {author}
  {\bibfnamefont {C.}~\bibnamefont {Sanchez}},\ }\href {\doibase
  10.1126/science.1232968} {\bibfield  {journal} {\bibinfo  {journal}
  {Science}\ }\textbf {\bibinfo {volume} {340}},\ \bibinfo {pages} {827}
  (\bibinfo {year} {2013})}\BibitemShut {NoStop}%
\bibitem [{\citenamefont {{Barker}}(2010)}]{2010PhRvL.105g3002B_barker_wgm}%
  \BibitemOpen
  \bibfield  {author} {\bibinfo {author} {\bibfnamefont {P.~F.}\ \bibnamefont
  {{Barker}}},\ }\href {\doibase 10.1103/PhysRevLett.105.073002} {\bibfield
  {journal} {\bibinfo  {journal} {\prl}\ }\textbf {\bibinfo {volume} {105}},\
  \bibinfo {eid} {073002} (\bibinfo {year} {2010})},\ \Eprint
  {http://arxiv.org/abs/1004.1443} {arXiv:1004.1443 [quant-ph]} \BibitemShut
  {NoStop}%
\bibitem [{\citenamefont {Garrett}\ \emph {et~al.}(2015)\citenamefont
  {Garrett}, \citenamefont {Somers},\ and\ \citenamefont
  {Munday}}]{Garrett:2015bde}%
  \BibitemOpen
  \bibfield  {author} {\bibinfo {author} {\bibfnamefont {J.~L.}\ \bibnamefont
  {Garrett}}, \bibinfo {author} {\bibfnamefont {D.}~\bibnamefont {Somers}}, \
  and\ \bibinfo {author} {\bibfnamefont {J.~N.}\ \bibnamefont {Munday}},\
  }\href {\doibase 10.1088/0953-8984/27/21/214012} {\bibfield  {journal}
  {\bibinfo  {journal} {J. Phys. Condens. Matter}\ }\textbf {\bibinfo {volume}
  {27}},\ \bibinfo {pages} {214012} (\bibinfo {year} {2015})},\ \Eprint
  {http://arxiv.org/abs/1409.5012} {arXiv:1409.5012 [cond-mat.other]}
  \BibitemShut {NoStop}%
\bibitem [{\citenamefont {Blakemore}\ \emph {et~al.}(2020)\citenamefont
  {Blakemore}, \citenamefont {Martin}, \citenamefont {Fieguth}, \citenamefont
  {Kawasaki}, \citenamefont {Priel}, \citenamefont {Rider},\ and\ \citenamefont
  {Gratta}}]{Blakemore2020}%
  \BibitemOpen
  \bibfield  {author} {\bibinfo {author} {\bibfnamefont {C.~P.}\ \bibnamefont
  {Blakemore}}, \bibinfo {author} {\bibfnamefont {D.}~\bibnamefont {Martin}},
  \bibinfo {author} {\bibfnamefont {A.}~\bibnamefont {Fieguth}}, \bibinfo
  {author} {\bibfnamefont {A.}~\bibnamefont {Kawasaki}}, \bibinfo {author}
  {\bibfnamefont {N.}~\bibnamefont {Priel}}, \bibinfo {author} {\bibfnamefont
  {A.~D.}\ \bibnamefont {Rider}}, \ and\ \bibinfo {author} {\bibfnamefont
  {G.}~\bibnamefont {Gratta}},\ }\href {\doibase 10.1116/1.5139638} {\bibfield
  {journal} {\bibinfo  {journal} {J. Vac. Sci. Tech. B}\ }\textbf {\bibinfo
  {volume} {38}},\ \bibinfo {pages} {024201} (\bibinfo {year}
  {2020})}\BibitemShut {NoStop}%
\bibitem [{\citenamefont {Dylla}\ and\ \citenamefont {King}(1973)}]{Dylla1973}%
  \BibitemOpen
  \bibfield  {author} {\bibinfo {author} {\bibfnamefont {H.~F.}\ \bibnamefont
  {Dylla}}\ and\ \bibinfo {author} {\bibfnamefont {J.~G.}\ \bibnamefont
  {King}},\ }\href {\doibase 10.1103/PhysRevA.7.1224} {\bibfield  {journal}
  {\bibinfo  {journal} {Phys. Rev. A}\ }\textbf {\bibinfo {volume} {7}},\
  \bibinfo {pages} {1224} (\bibinfo {year} {1973})}\BibitemShut {NoStop}%
\bibitem [{\citenamefont {Marocco}\ and\ \citenamefont
  {Sarkar}(2021)}]{Marocco2021}%
  \BibitemOpen
  \bibfield  {author} {\bibinfo {author} {\bibfnamefont {G.}~\bibnamefont
  {Marocco}}\ and\ \bibinfo {author} {\bibfnamefont {S.}~\bibnamefont
  {Sarkar}},\ }\href {\doibase 10.21468/SciPostPhys.10.2.043} {\bibfield
  {journal} {\bibinfo  {journal} {SciPost Phys.}\ }\textbf {\bibinfo {volume}
  {10}},\ \bibinfo {pages} {43} (\bibinfo {year} {2021})}\BibitemShut {NoStop}%
\bibitem [{\citenamefont {Dubovsky}\ \emph {et~al.}(2004)\citenamefont
  {Dubovsky}, \citenamefont {Gorbunov},\ and\ \citenamefont
  {Rubtsov}}]{Dubovsky2004WMAP}%
  \BibitemOpen
  \bibfield  {author} {\bibinfo {author} {\bibfnamefont {S.}~\bibnamefont
  {Dubovsky}}, \bibinfo {author} {\bibfnamefont {D.~S.}\ \bibnamefont
  {Gorbunov}}, \ and\ \bibinfo {author} {\bibfnamefont {G.~I.}\ \bibnamefont
  {Rubtsov}},\ }\href@noop {} {\bibfield  {journal} {\bibinfo  {journal} {J.
  Exp. Theor. Phys. Lett.}\ }\textbf {\bibinfo {volume} {79}},\ \bibinfo
  {pages} {1} (\bibinfo {year} {2004})}\BibitemShut {NoStop}%
\bibitem [{\citenamefont {Dolgov}\ \emph {et~al.}(2013)\citenamefont {Dolgov},
  \citenamefont {Dubovsky}, \citenamefont {Rubtsov},\ and\ \citenamefont
  {Tkachev}}]{Dolgov:2013una}%
  \BibitemOpen
  \bibfield  {author} {\bibinfo {author} {\bibfnamefont {A.~D.}\ \bibnamefont
  {Dolgov}}, \bibinfo {author} {\bibfnamefont {S.~L.}\ \bibnamefont
  {Dubovsky}}, \bibinfo {author} {\bibfnamefont {G.~I.}\ \bibnamefont
  {Rubtsov}}, \ and\ \bibinfo {author} {\bibfnamefont {I.~I.}\ \bibnamefont
  {Tkachev}},\ }\href {\doibase 10.1103/PhysRevD.88.117701} {\bibfield
  {journal} {\bibinfo  {journal} {Phys. Rev. D}\ }\textbf {\bibinfo {volume}
  {88}},\ \bibinfo {pages} {117701} (\bibinfo {year} {2013})}\BibitemShut
  {NoStop}%
\bibitem [{\citenamefont {Boddy}\ \emph {et~al.}(2018)\citenamefont {Boddy},
  \citenamefont {Gluscevic}, \citenamefont {Poulin}, \citenamefont {Kovetz},
  \citenamefont {Kamionkowski},\ and\ \citenamefont {Barkana}}]{Boddy:2018wzy}%
  \BibitemOpen
  \bibfield  {author} {\bibinfo {author} {\bibfnamefont {K.~K.}\ \bibnamefont
  {Boddy}}, \bibinfo {author} {\bibfnamefont {V.}~\bibnamefont {Gluscevic}},
  \bibinfo {author} {\bibfnamefont {V.}~\bibnamefont {Poulin}}, \bibinfo
  {author} {\bibfnamefont {E.~D.}\ \bibnamefont {Kovetz}}, \bibinfo {author}
  {\bibfnamefont {M.}~\bibnamefont {Kamionkowski}}, \ and\ \bibinfo {author}
  {\bibfnamefont {R.}~\bibnamefont {Barkana}},\ }\href {\doibase
  10.1103/PhysRevD.98.123506} {\bibfield  {journal} {\bibinfo  {journal} {Phys.
  Rev. D}\ }\textbf {\bibinfo {volume} {98}},\ \bibinfo {pages} {123506}
  (\bibinfo {year} {2018})}\BibitemShut {NoStop}%
\end{thebibliography}%

\end{document}